\documentclass[aps,prl,reprint]{revtex4-2}

\usepackage{amsmath}
\usepackage{amssymb}
\usepackage{graphicx}
\usepackage{dcolumn}
\usepackage{bm}
\usepackage{multirow}
\usepackage{makecell}
\usepackage{booktabs}
\usepackage{gensymb}
\usepackage{array}

\usepackage[T1]{fontenc}
\graphicspath{{Figures/}}

\begin{document}

\title{Modulated Anti-Ferroelectric Smectic Phases with Orthogonal and Tilted Structures}

\author{Jordan Hobbs$^1$}
\author{Calum J. Gibb$^2$}
\author{William C. Ogle$^2$}
\author{Peter Medle Rupnik$^{3,4}$}
\author{Natan Osterman$^{3,4}$}
\author{Nerea Sebastián$^3$}
\author{Alenka Mertelj$^3$}
\author{Richard. J. Mandle$^{1,2}$}
 \email{r.j.mandle@leeds.ac.uk}
\affiliation{$^1$School of Physics \& Astronomy, University of Leeds, Leeds, UK}
\affiliation{$^2$School of Chemistry, University of Leeds, Leeds, UK}
\affiliation{$^3$Complex Matter Department, Jožef Stefan Institute, Ljubljana, Slovenia}
\affiliation{$^4$Faculty of Mathematics and Physics, University of Ljubljana, Ljubljana, Slovenia}

\date{\today}

\begin{abstract}
The discovery of the ferroelectric nematic phase has brought with it a plethora of new polar liquid crystalline phases. One in particular is the anti-ferroelectric smectic A SmA\textsubscript{AF} phase. In this letter we show via observation and analysis of satellite peaks in the X-ray scattering patten that the structure of the SmA\textsubscript{AF} phase involves a density modulation of $\approx$10-20 nm lateral to the smectic layer normal. Further, we demonstrate a previously undiscovered phase where the anti-ferroelectric order is maintained into a tilted smectic phase demonstrating the robustness of the underlying frustration that leads to the modulated structure. We suggest that the modulations are only in a single dimension and appear parallel to the tilt plane. This new phase also shows a significantly different and complex response to an electric field from other discovered polar LC phases due to the ability to modulate both tilt and polarisation direction. 
\end{abstract}

\maketitle

The discovery of the ferroelectric nematic phase (N\textsubscript{F}) in 2017 \cite{2017_Mandle_1, 2017_Nishikawa, 2020_Chen, 2020_Sebastian} has been accompanied by the the subsequent discovery of a flurry of associated phases including positionally ordered orthogonal \cite{2022_Kikuchi, 2022_Chen2} and tilted \cite{2024_Kikuchi, 2025_Hobbs1, 2025_Strachan} ferroelectric  and anti-ferroelectric phases \cite{2018_Mertelj, 2020_Sebastian, 2023_Chen, 2024_Gibb, 2024_Hobbs}. The anti-ferroelectric splay nematic (N\textsubscript{S}) phase (sometimes referred to as the SmZ\textsubscript{A} phase) exhibits nematic orientational order of the constituent molecules with an average director and either a 1D \cite{2023_Chen, 2024_Nacke, 2024_Thapa} or 2D \cite{2020_Rosseto, 2025_Rupnik, 2025_Ma} density modulation perpendicular to the director. The modulations consist of alternating domains of polarisation splay separated by Ising walls. Across the domains the density modulates, probably sinusoidally \cite{2023_Chen}, such that it is maximised at the centre of the splay domains and minimized at the domain walls which consist of apolar nematic regions with reduced mass density \cite{2024_Nacke} which can result in weak but observable Bragg scattering peak \cite{2023_Chen}. These modulated structures appear due to competition between flexoelectric effects and electrostatics \cite{2025_Rupnik}.

The smectic equivalent of the N\textsubscript{S} phase is the anti-ferroelectric smectic A (SmA\textsubscript{AF}). This phase has been observed in single component materials \cite{2024_Gibb} as well as mixtures \cite{2024_Hobbs, 2025_Pociecha} and has been the subject of theoretical analysis \cite{2026_Mukherjee} but has so far been assumed to be the smectic analogue of the N\textsubscript{S} phase i.e. alternating splay domains but with the additional component of positional order in the form of smectic layers with their planes normal to the director. However, there has been no experimental verification of this and there are also significant open questions such as whether the phase exhibits 1D or 2D splay modulation \cite{2020_Rosseto, 2025_Rupnik}. 

In this letter we demonstrate conclusively that the SmA\textsubscript{AF} phase exhibits an additional density modulation perpendicular to the layer planes due to the 1D splay domains. We also show that these modulations continue to exist even when molecules begin to tilt away from the layer normal in the form of a newly discovered anti-ferroelectric smectic C (SmC\textsubscript{AF}) phase which also exhibits 1D splay.

\begin{table*}
\centering
\caption{Transition temperatures (T) and associated enthalpies of transition ($\Delta$H) for compounds 1–2 determined by DSC on cooling at a rate of 10 °C min$^{-1}$. A $^m$ indicates determination via POM observations. Both materials exhibit signs of slight degradation in their DSC obtained transition temps and so these are consistently 1-3$^\circ$C lower than when measured by other transition techniques.}
\label{tbl:structures}
\resizebox{\textwidth}{!}{%
\begin{tabular}{cccccccccc}\toprule
 No. & & &
  \textbf{Melt} &
  \textbf{SmC$_\text{P}^\text{H}$-SmC$_\text{AF}$} &
  \textbf{SmC$_\text{AF}$-SmA$_\text{AF}$} &
  \textbf{SmA$_\text{AF}$-SmA} &
  \textbf{SmA-N} &
  \textbf{N-Iso} \\ \midrule
\textbf{1}& \raisebox{-.5\height}{\includegraphics[width=0.3\linewidth]{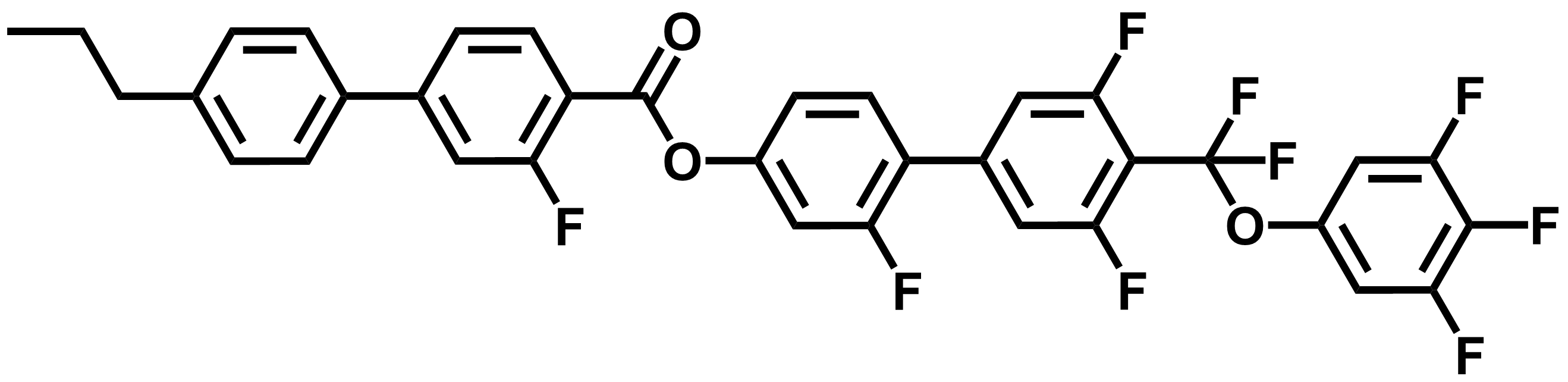}} & \makecell{T (°C)\\$\Delta$H (kJ/mol)} & \makecell{117.9\\23.6} & \makecell{120.6\\0.3} & \makecell{122.7\\0.6} & \makecell{133.8\\0.05} & \makecell{228$^m$\\-} & \makecell{273.2$^m$\\-} \\
\textbf{2}& \raisebox{-.33\height}{\includegraphics[width=0.25\linewidth]{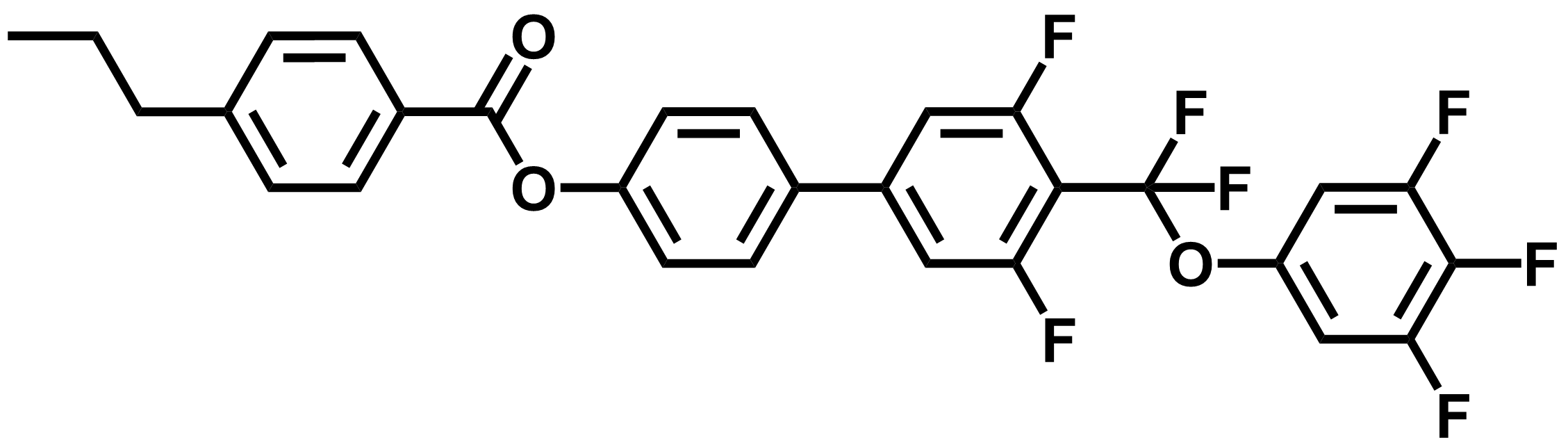}} & \makecell{T (°C)\\$\Delta$H (kJ/mol)} & \makecell{92.3\\21.7} & - & \makecell{86.6\\0.08} & \makecell{98.2\\0.04} & \makecell{127.7\\0.3} & \makecell{171.4\\0.9} \\ \bottomrule
\end{tabular}%
}
\end{table*}

Compound \textbf{1} exhibits five mesophases with the phase order of N-SmA-SmA\textsubscript{AF}-SmC\textsubscript{AF}-SmC$_\text{P}^\text{H}$ with the transition temperatures shown in Table \ref{tbl:structures}. The transition from the SmA-SmA\textsubscript{AF} phase generally shows very little textural change \cite{2024_Gibb}, with the observations here consistent with that observation. In planar aligned cells the SmA phase exhibits parabolic focal conics \cite{1977_Rosenblatt, 1989_Ouchi}, which are more prominent in thicker cells (Fig. S2). These defects are retained upon cooling into the SmA\textsubscript{AF} phase albeit they become significantly more visible. Upon further cooling into the SmC\textsubscript{AF} phase the defect segment within the parabola splits down the middle into a darker and light side. The extinct side can be switched by rotating the sample past the analyser indicating the change in director orientation within each subdomain of the defect (fig. S3). In the SmC$_\text{P}^\text{H}$ periodic stripes appear perpendicular to the rubbing direction. The periodicity of these stripes depends only weakly on temperature, with a slight reduction in periodicity with reducing temperature, but are independent of cell thickness. Light diffraction measurements reveal a periodicity of 2 $\mu$m at 116$^\circ$C. Diffraction measurements above this temperature did not show a clear signal preventing measurement of the exact temperature dependence of the periodicity.

\begin{figure*}
    \centering
    \includegraphics[width=\linewidth]{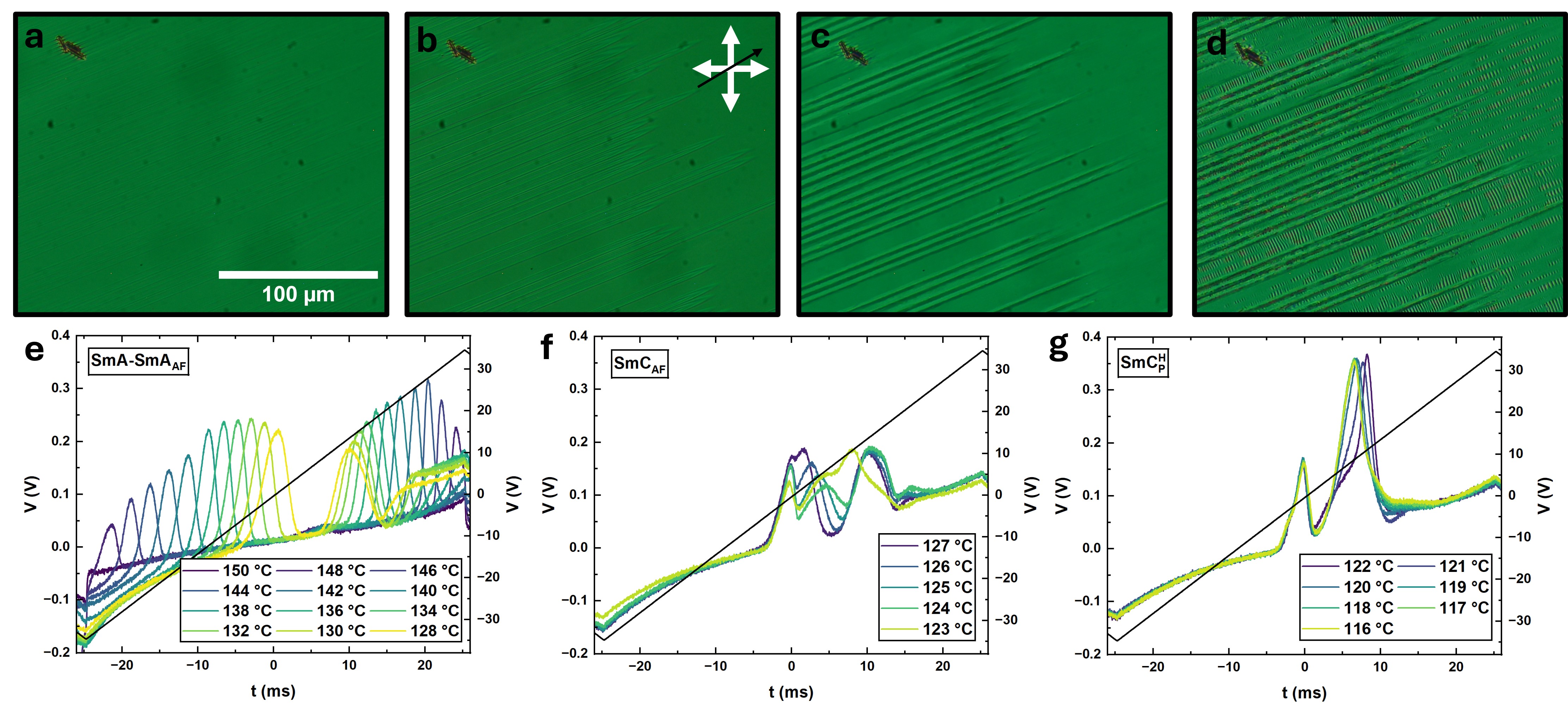}
    \caption{POM images of compound \textbf{1} in a) the SmA, b) the SmA\textsubscript{AF}, c) the SmC\textsubscript{AF} and d) the SmC$_\text{P}^\text{H}$ phases. POM images were taken in a 5 $\mu$m cell with syn-parallel rubbed planar aligned substrates. White arrows indicate polarisor and analyser orientations and black arrow the rubbing direction. Current response measurements for compound \textbf{1} in a 5 $\mu$m cell with no alignment layer taken with a 10 Hz, 20 V\textsubscript{RMS} triangular voltage profile e) over the SmA-SmA\textsubscript{AF} phases, f) the SmC\textsubscript{AF} and g) the SmC$_\text{P}^\text{H}$ phase.}
    \label{fig:CJG333POM}
\end{figure*}

Current response measurements demonstrate the polar nature of the various mesophases. The SmA\textsubscript{AF} phase shows double peaks in the current response indicating the presence of the two anti-ferroelectric sublattice domains (Fig. \ref{fig:CJG333POM}e). The double peaks are also observed in the preceding SmA phase due to a high temperature critical end point \cite{2023_Szydlowska}. As the temperature is reduced the peaks move closer to t = 0 s. At the transition to the SmC\textsubscript{AF} phase a third peak emerges (Fig. \ref{fig:CJG333POM}f), moving out of the negative polarity peak and to longer timescales with a similar temperature dependence as for the negative polarity peak in the preceding SmA\textsubscript{AF} phase. Eventually the peak merges with the positive polarity peak at the transition to the SmC$_\text{P}^\text{H}$ phase. In the SmC$_\text{P}^\text{H}$ phase (Fig. \ref{fig:CJG333POM}g), the familiar small peak pre-voltage polarity reversal corresponding to tilt reformation and a larger peak due to simultaneous (or close to) tilt removal and polarisation reversal \cite{2025_Hobbs2}. 

Second harmonic generation (SHG) measurements show only a very weak signal in the SmA\textsubscript{AF} and SmC\textsubscript{AF} phases where SHG microscopy (Fig. S7) reveals that the origin of this signal is non-uniform and generally localised to defects thus we take this data as supportive of our phase assignment. The SmC$_\text{P}^\text{H}$ phase exhibits an SHG signal more than 5 orders of magnitude larger than the SmC\textsubscript{AF} phase (Fig. S7), consistent with the suggested structure and previous measurements \cite{2024_Gibb}.

\begin{figure*}
    \centering
    \includegraphics[width=\linewidth]{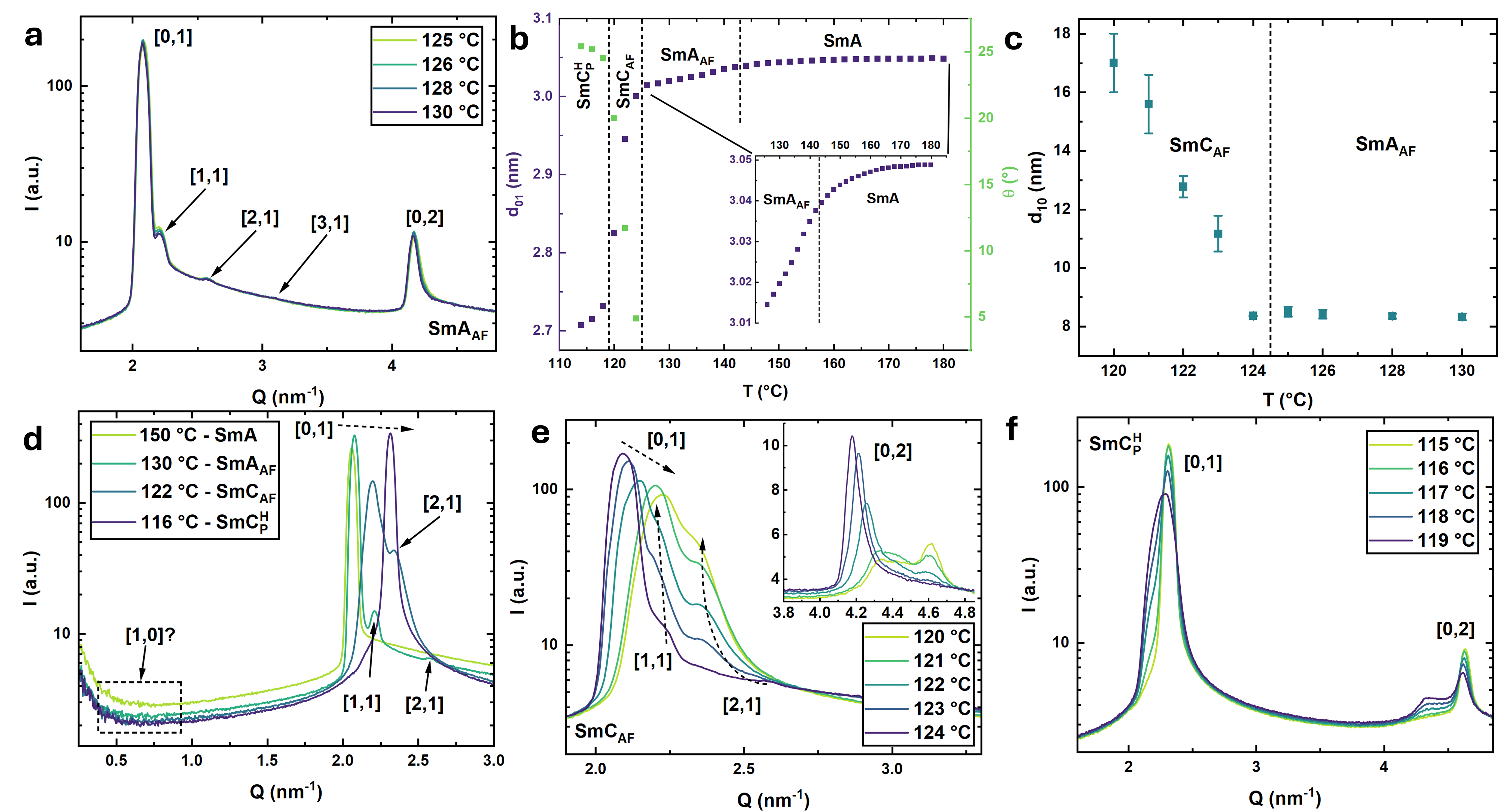}
    \caption{a) 1D X-ray scattering pattern for compound \textbf{1} in the SmA\textsubscript{AF} phase. b) Temperature dependence of the layer thickness with the inset showing a zoomed in region of the SmA and SmA\textsubscript{AF} phase. c) Temperature dependence of the lateral density modulation thickness. d) 1D X-ray scattering pattern with a different sample-detector distance showing the absence of a [1,0] peak for all phases. 1D X-ray scattering patterns for the e) SmC\textsubscript{AF} phase and the f) SmC$_\text{P}^\text{H}$ phase.}
    \label{fig:CJG333_Xay}
\end{figure*}

X-ray scattering measurements in the SmA phase show a single Bragg peak resulting from the smectic layers. Upon transition to the SmA\textsubscript{AF} phase an additional scattering peak emerges at higher Q-spacing than the layer spacing peak. As the temperature is reduced, the intensity of the additional peak increases followed by the appearance of two further peaks at even higher Q-spacing (Fig. \ref{fig:CJG333_Xay}a). The X-ray scattering pattern can be well indexed by a 2D rectangular lattice where the layer peak is the [0,1] peak and the others are successive [h,1] peaks both considering their Q-space and azimuthal positions (Fig. S6). While the d\textsubscript{01} length-scale (layer separation) is weakly temperature dependent, decreasing slightly as the temperature is reduced (Fig. \ref{fig:CJG333_Xay}b), the d\textsubscript{10} distance implied by the indexing is temperature independent with a length-scale of 8.4 nm (Fig. \ref{fig:CJG333_Xay}c). For the data presented here, no peak in that region could be observed (Fig. \ref{fig:CJG333_Xay}d). This is perhaps not surprising as all observations of the Bragg peak associated with this lateral density modulation for the N\textsubscript{S} phase published so far \cite{2022_Nishikawa, 2023_Chen, 2023_Cruickshank} have been obtained using synchrotron radiation while this data is collected using a lab based scattering setup with around 5 orders of magnitude less flux. The Bragg peak in the N\textsubscript{S} phase is extremely weak due to the low amplitude of the associated density modulation and this is consistent with the data presented here for the SmA\textsubscript{AF} phase, although the presence of higher order peaks may imply that the lateral density modulation is longer range than the smectic layers despite the lower amplitude of the modulation. For the N\textsubscript{S} phase, satellite peaks have not been observed as the [0,1] peak is not Bragg like (the phase is nematic) and is broader and less intense thus any satellite peaks would be orders of magnitude weaker and hidden by the broad [0,1] peak.

The SmC\textsubscript{AF} phase can be indexed using the same 2D rectangular lattice as the SmA\textsubscript{AF} (Fig. \ref{fig:CJG333_Xay}e). Non-orthogonal lattices such as 2D oblique or 3D monoclinic patterns would show peaks with negative indices at different Q values (e.g. both [1,1] and [-1,1]) and these are not observed. The [0,1] peak moves to longer Q as the molecules in the layers tilt away from the layer normal leading to a reduction in the layer spacing, d\textsubscript{01}. For most polar smectics the tilt develops over 10-20$^\circ$C before saturating \cite{2025_Hobbs1, 2024_Kikuchi, 2025_Hobbs2} but here it occurs rapidly, reaching $\approx$ 20$^\circ$ within 4$^\circ$C (Fig. \ref{fig:CJG333_Xay}b). The [n,1] peaks move towards the [0,1] peak indicating an increase in the lateral density modulation period, d\textsubscript{10}. Upon transition to the SmC$_\text{P}^\text{H}$ phase the satellite peaks vanish and only a single [0,1] peak is left (Fig. \ref{fig:CJG333_Xay}f) indicating a melting of the lateral density modulation at the transition from the anti-ferroelectric SmC\textsubscript{AF} phase to the ferroelectric SmC$_\text{P}^\text{H}$ phase.

\begin{figure}
    \centering
    \includegraphics[width=1\linewidth]{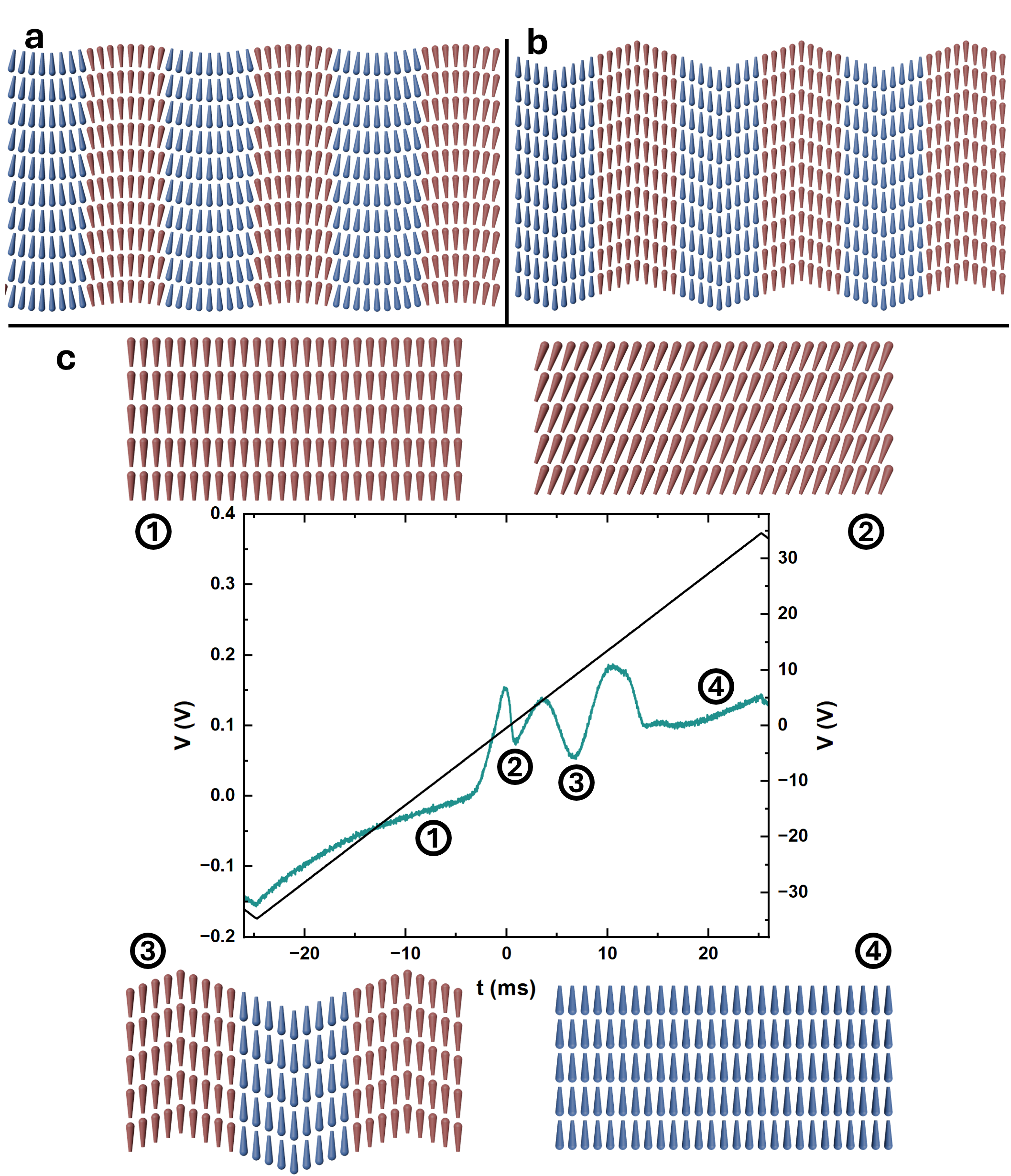}
    \caption{Schematics of the proposed structure of the a) SmA\textsubscript{AF} and b) SmC\textsubscript{AF} phase. Regions of alternate polarisation are separated by Ising walls while in the SmC\textsubscript{AF} splay is limited to the regions mediating change of tilt sign. While the SmC\textsubscript{AF} schematic has been drawn fully chevron, in reality this structure is likely weaker and more sinusoidal. c) Current response measurement for compound \textbf{1} in the SmC\textsubscript{AF} phase in a 5 $\mu$m thick cell with no alignment layer. (1) and (4) show the saturated SmA\textsubscript{F} state, (2) shows the SmC\textsubscript{P} state and (3) shows the ground state SmC\textsubscript{AF} phase.}
    \label{fig:CJG333_Switching}
\end{figure}

For the N\textsubscript{S} phase both 1D \cite{2023_Chen} and 2D \cite{2025_Rupnik, 2025_Ma} have been proposed and so we consider both options for the SmA\textsubscript{AF} phase here. A 2D splayed SmA\textsubscript{AF} should give additional scattering peaks (for example a [1,1,1] peak using a tetragonal lattice) however these additional cross scattering peaks are not observed which strongly suggests 1D splay rather than 2D. Additionally, the very weak SHG signal measured in the SmA\textsubscript{AF} and SmC\textsubscript{AF} phases is supportive of 1D splay as the origin of the signal is likely irregularities and defects. In a 2D splayed system with modulation period in 10 nm range these would cancel out but in a 1D the cancellation might be lost around defects and a weak signal could be observed. The proposed SmA\textsubscript{AF} structure is shown in fig. \ref{fig:CJG333_Switching}a.

For the SmC\textsubscript{AF} the possibilities are wider as 1D anti-ferroelectric modulations could be perpendicular, parallel or at some angle to the tilt plane while 2D modulations would be a combination of these possibilities. The X-ray is again only fitted using a 2D rectangular lattice suggesting 1D splay while POM observations (Fig. S3) strongly suggest that the structure must be optically uniaxial. From these two observations we suggest that the structure of the SmC\textsubscript{AF} phase presented here is of alternating polarisation domains comprised of alternating positive and negative tilt sub-domains. The alternating tilt sub-domains of same polarisation sign are joined by a regions of splay while the opposite polarisation domains are joined by Ising walls. The proposed structure is shown in fig. \ref{fig:CJG333_Switching}b.

Having established the ground state structure of the SmC\textsubscript{AF} phase, the effects of electric field on the structure can be presented (Fig. \ref{fig:CJG333_Switching}c). Since other longitudinally polar mesophases have shown the ability to undergo a field induced phase transition to their orthogonal (non-tilted) equivalent \cite{2025_Hobbs1, 2025_Hobbs2, 2025_Strachan, 2025_Gorecka} a sufficiently high positive or negative voltage results in a field induced phase transition to a SmA\textsubscript{F} state (Fig. \ref{fig:CJG333_Switching}c(1)) as the saturated state of electric field application. Increasing the voltage towards 0 V allows the tilt to reform giving a SmC\textsubscript{P} state (Fig. \ref{fig:CJG333_Switching}c(2)). Further increase of the voltage towards 0 V and beyond results in the ground state of the SmC\textsubscript{AF} phase (Fig. \ref{fig:CJG333_Switching}c(3)). This happens past the 0 V position due to the viscosity and dynamics of the system meaning it lags behind the electric field switching. Increasing the voltage towards the saturation voltage results in a field induced phase transition to a SmA\textsubscript{F} with opposite polarisation to the initial state (Fig. \ref{fig:CJG333_Switching}c(4)).

\begin{figure}
    \centering
    \includegraphics[width=0.9\linewidth]{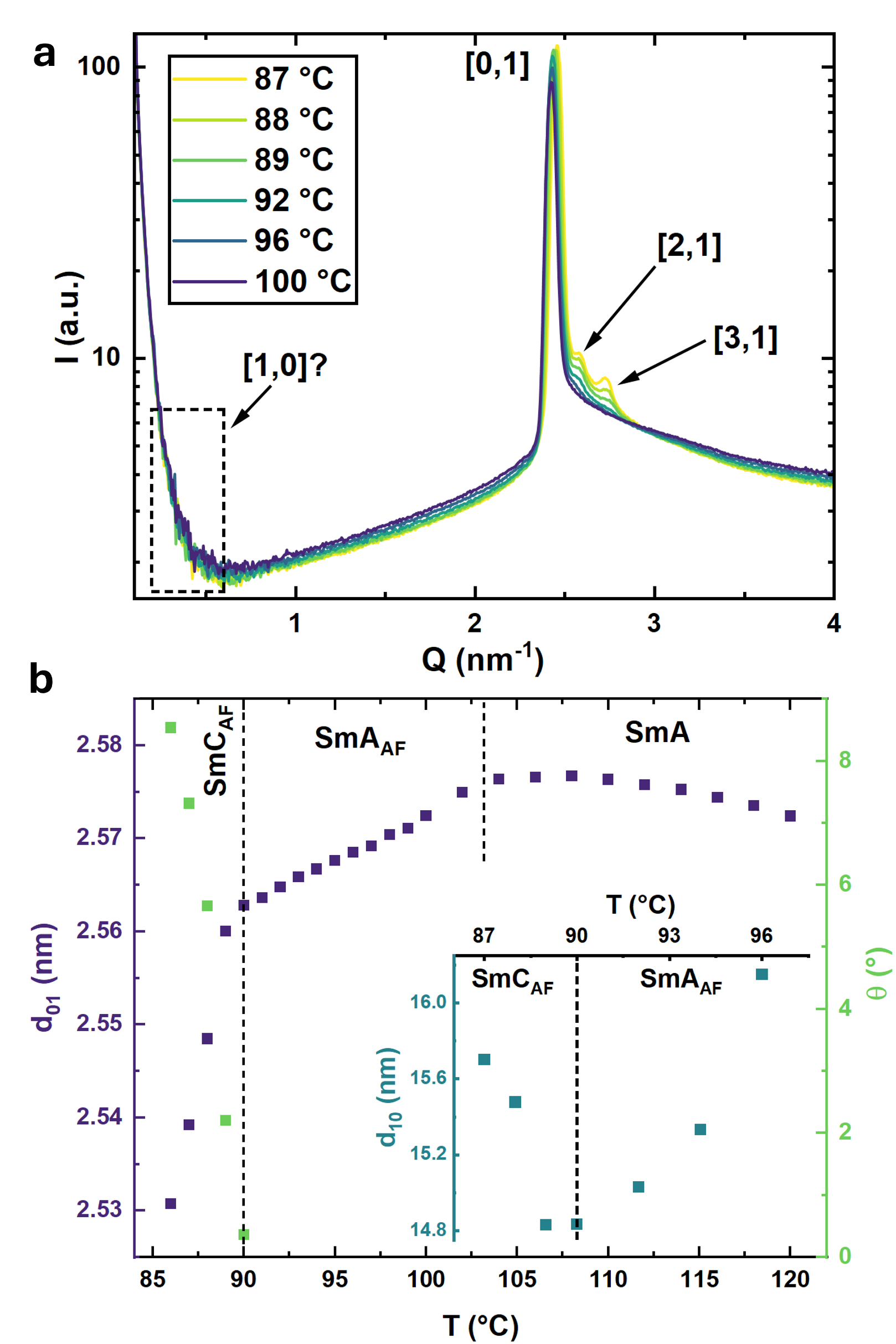}
    \caption{a) 1D X-ray scattering pattern for compound \textbf{2} throughout both the SmA\textsubscript{AF} and SmC\textsubscript{AF} phase with the transition between the two at 90$^\circ$C. b) Temperature dependence of the smectic layer thickness,d\textsubscript{01}, with the inset showing the dependence of the lateral density modulation, d\textsubscript{10}.}
    \label{fig:CJG123_Xray}
\end{figure}

Compound \textbf{2} has previously been reported to exhibit N-SmA-SmA\textsubscript{AF}-SmC$_\text{P}^\text{H}$ phase behaviour \cite{2024_Gibb}, although no satellite peaks where observed in the SmA\textsubscript{AF} phase so no verification of the structure was presented. However, if the data collection time is increased (here by a factor of 10) then weak peaks can be observed in both the SmA\textsubscript{AF} and the reported SmC$_\text{P}^\text{H}$ phase (Fig. \ref{fig:CJG123_Xray}a). Neither the SmA\textsubscript{AF} or SmC$_\text{P}^\text{H}$ phase exhibit any SHG signal \cite{2024_Gibb} and so we conclude that the SmC$_\text{P}^\text{H}$ phase reported for compound \textbf{2} previously is actually another example of the SmC\textsubscript{AF} phase. When the two additional peaks observed in both the SmA\textsubscript{AF} and SmC\textsubscript{AF} phase are indexed, we find that the [1,1] peak is hidden by the intense [0,1] peak associated with the smectic layers indicating a much longer pitch of the lateral modulation than for compound \textbf{1}. The satellite peak indexing predict a consistent position for the [1,0] peak although no peak is observed due to the apparent weakness of the density modulation amplitude. For compound \textbf{2} the SmA\textsubscript{AF} shows a reduction in the lateral density modulation throughout the phase before it then increases in the SmC\textsubscript{AF} phase as for in compound \textbf{1} (Fig. \ref{fig:CJG123_Xray}b). The increased d\textsubscript{01} of compound \textbf{2} vs compound \textbf{1} explains why these peaks were not observed previously as it shifts all the satellite peaks closer to the to intense [0,1] peak hiding the most intense and most obvious peak in the main [0,1] peak. Once again we see that the transition between the SmA\textsubscript{AF} and SmC\textsubscript{AF} phase shows a continuation of the same lattice and we suggest that here again both materials show 1D modulations with the modulations parallel to the tilt plane in the SmC\textsubscript{AF} phase.

In conclusion, it was shown that the SmA\textsubscript{AF} phase is modulated and exhibits a 1D splay modulation rather than 2D as well as the discovery of a new polar liquid crystalline phase, the SmC\textsubscript{AF} phase in compounds \textbf{1} and \textbf{2} which exhibits a 1D chevron modulated structure, parallel to the tilt plane. The appearance of these modulated smectic phases demonstrates the strength of the flexoelectric coupling found in these polar systems as they are strong enough to result in the deformation of not just nematic phases \cite{2018_Mertelj, 2020_Sebastian, 2020_Chen} but now conclusively in both SmA and SmC type phase structures where the free energy cost of the deformation is presumably increased due to the stiffness of the layers and the inability to accommodate bend deformations. Additional references have been cited in the Supplementary Information \cite{2023_Sebastian}.

\section{Data availability}

The data associated with this paper is openly available from the University of Leeds Data Repository at: https://doi.org/10.5518/1806

\section{Acknowledgements}

R.J.M. thanks UKRI for funding via a Future Leaders Fellowship, grant number MR/W006391/1, and the University of Leeds for funding via a University Academic Fellowship. R.J.M. and J.H. thank the Royal Society for funding via Research Grant RGS/R2/242503. R.J.M and W.C.O thank the EPSRC CDT in Soft Matter for Formulation and Industrial Innovation (SOFI2), (EP/S023631/1). R.J.M. gratefully acknowledges support from Merck KGaA. N.S. and A.M. acknowledge the support of the Slovenian Research Agency (Grant Nos. P1-0192, J1-50004, and BI-VB/25-27-011). The authors acknowledge EPSRC for funding the SAXS/WAXS system via a capital equipment grant EP/X0348011.

\section{Author Contribution Statement}

J.H conducted POM, DSC, current response and X-ray scattering data. C.J.G and W.C.O completed the chemical synthesis and structural characterisation. P.M.R, N.O, N.S, and A.M conducted SHG measurements. All authors contributed to interpretation of all data and the final manuscript.

\section{Competing Interests}
Authors declare that they have no competing interests

\bibliography{bib}

\end{document}


\maketitle

\section{Methods}

\subsection{Chemical Synthesis}
Chemicals were purchased from commercial suppliers (Fluorochem, Merck, Apollo Scientific) and used as received. Solvents were purchased from Merck and used without further purification. Reactions were performed in standard laboratory glassware at ambient temperature and atmosphere and were monitored by TLC with an appropriate eluent and visualised with 254 nm or 365 nm light. Chromatographic purification was performed using a Combiflash NextGen 300+ System (Teledyne Isco) with a silica gel stationary phase and a hexane/ethyl acetate gradient as the mobile phase, with detection made in the 200-800 nm range. Chromatographed materials subjected to re-crystallisation from an appropriate solvent system.

\subsection{Chemical Characterisation Methods}
The structures of intermediates and final products were determined using \ce{^1H}, \ce{^{13}C\{^1H\}}, and \ce{^{19}F} NMR spectroscopy. NMR spectroscopy was performed using either a Bruker Avance III HDNMR spectrometer operating at 400 MHz, 100.5 MHz or 376.4 MHz (\ce{^1H}, \ce{^{13}C\{^1H\}} and \ce{^{19}F}, respectively) or a Bruker AV4 NEO 11.75T spectrometer operating at 500 MHz, 125.5 MHz (\ce{^1H} and \ce{^{13}C\{^1H\}}, respectively). Unless otherwise stated, spectra were acquired as solutions in deuterated chloroform, coupling constants are quoted in Hz, and chemical shifts are quoted in ppm.
\subsection{Thermal Analysis}

Differential scanning calorimetry (DSC) measurements were performed using a TA Instruments Q2000 DSC instrument (TA Instruments, Wilmslow UK), equipped with a RCS90 Refrigerated cooling system (TA Instruments, Wilmslow UK). The instrument was calibrated against an Indium standard, and data were processed using TA Instruments Universal Analysis Software. Samples were analysed under a nitrogen atmosphere, in hermetically sealed aluminium TZero crucibles (TA Instruments, Wilmslow, UK) and subjected to three analysis cycles. In all cases, samples were subject to heating and cooling at a rate of 10 K/min.  Phase transition temperatures were measured as onset values on cooling cycles for consistency between monotropic and enantiotropic phase transitions, while crystal melts were obtained as onset values on heating. 

\subsection{Optical Microscopy}

Polarised light optical microscopy (POM) was performed using a Leica DM2700P polarised light microscope (Leica Microsystems (UK) Ltd., Milton Keynes, UK). A Linkam TMS 92 heating stage (Linkam Scientific Instruments Ltd., Redhill, UK) was used for temperature control. Images were recorded using a Nikon D3500 Digital Camera (Nikon UK Ltd., Surbiton, UK), using DigiCamControl software. Samples were studied sandwiched between two untreated glass coverslips or in cells purchased from Military University of Technology, Poland.

\subsection{X-ray Scattering}

X-ray scattering measurements, both small angle (SAXS) and wide angle (WAXS) were recorded using an Anton Paar SAXSpoint 5.0. This was equipped with a primux 100 Cu X-ray source with a 2D EIGER2 R detector with a variable sample-detector distance. The X-rays had a wavelength of 0.154 nm. Samples were filled into either thin-walled quartz capillaries or held between Kapton tape. Temperature was controlled using an Anton Paar heated sampler with a range of 20$^\circ$C to 300$^\circ$C and the samples held in a chamber with an atmospheric pressure of $>$1 mbar.

Background scattering patterns were recorded, scaled according to the sample transmission and then subtracted from the samples' obtained 2D SAXS pattern. 1D patterns were obtained by radially integrating the 2D SAXS patterns. Peak positions and FWHM was recorded and then converted into d spacing following Bragg's law. In tilted smectic phases, the tilt was obtained from:
\begin{equation}
    \frac{d_{c}}{d_{A}} = \cos\theta
\end{equation}
where $d_c$ is the layer spacing in the tilted smectic phase, $d_A$ is the extrapolated spacing from the non-tilted preceding phase, extrapolated to account for the weak temperature dependence of the preceding phases due to shifts in conformation and order, and $\theta$ the structural tilt angle.

For the modulated structures here the $d_{hk}$ spacing of each of the satellite peaks, [h,k], can be calculated from:
\begin{equation}
    \frac{1}{d_{hk}^2}=\frac{h^2}{d_{10}^2}+\frac{k^2}{d_{01}^2}
\end{equation}
while the azimuthal angular position, $\phi_{hk}$, of a satellite peak is thus:
\begin{equation}
    \phi_{hk}=\tan^{-1}\left (\frac{kd_{10}}{hd_{01}}\right )
\end{equation}
i.e. $d_{10}$ is at 90$^\circ$ and $d_{01}$ is at 0$^\circ$. In reality the [1,0] peak is not observed so the index of each satellite was systematically guessed until they produced consistent predictions for the [1,0] peak position. The position of the [1,0] peak was taken as the average of the predicted positions.

\subsection{Current Response Measurements}

Current Response measurements are undertaken using the current reversal technique. Triangular waveform AC voltages are applied to the sample cells with an Agilent 33220A signal generator (Keysight Technologies), and the resulting current outflow is passed through a current-to-voltage amplifier and recorded on a RIGOL DHO4204 high-resolution oscilloscope (Telonic Instruments Ltd, UK). Heating and cooling of the samples during these measurements is achieved with an Instec HCS402 hot stage controlled to 10 mK stability by an Instec mK1000 temperature controller. The LC samples are held in 4µm thick cells with no alignment layer, supplied by Instec. The measurements consist of cooling the sample at a rate of 1 Kmin\textsuperscript{-1} and applying a set voltage at a frequency of 20 Hz to the sample every 1 K. The voltage was set such that it would saturate the measured P\textsubscript{S} and was determined before final data collection.

There are three contributions to the measured current trace: accumulation of charge in the cell (I\textsubscript{c}), ion flow (I\textsubscript{i}), and the current flow due to polarisation reversal (I\textsubscript{p}). To obtain a P\textsubscript{Sat} value, we extract the latter, which manifests as one or multiple peaks in the current flow, and integrate as:
\begin{equation}
    P_{Sat} = \int_{}^{}{\frac{I_{p}}{2A}dt}
\end{equation}
where $A$ is the active electrode area of the sample cell.

\subsection{Pitch Measurements}

The pitch was measured by illumination of the sample from the bottom via a 405 nm laser. Since the pitch axis of the bulk will fall along the director, thus planar aligned cells were used to measure the pitch via light diffraction. The angle of diffraction was measured on a flat screen set away from the sample.

\subsection{SHG microscopy}
SHG investigations were performed using EHC D-type 5$\mu$m and 10$\mu$m thick parallel rubbed cells. SHG microscopy imaging is performed using a custom-built sample-scanning microscope. A detailed description of the setup can be found in reference \cite{2023_Sebastian}. The laser source is an Erbium-doped fiber laser (C-Fiber A 780, MenloSystems) generating 785 nm, 95 fs pulses at a 100 MHz repetition rate. A combination of galvo mirrors and a long-working distance objective (Nikon CFI T Plan SLWD, NA 0.3) is used to scan the focused beam in the sample plane. The scanning frequencies are much higher (a few 100 Hz) than the imaging frame rate (a few Hz). A long-working distance 20× objective (Nikon CFI T Plan SLWD, NA 0.3) collects the light coming from the sample. A set of 700 nm short-pass and 400 nm band-pass filters eliminates the fundamental IR light and any possible fluorescence signal. The images are finally acquired using a high-performance CMOS camera (Grasshopper 3, Teledyne Flir) with a typical integration time of 250 ms, dimensions 1920 × 1200 pixels, 12-bit depth, and 0.285 $\mu$m/pixel. A motorized half-waveplate for 800 nm rotates the polarization of the fundamental IR beam in the plane and, jointly with the analyser in front of the camera, enables to perform polarization resolved SHG. In the case of SHG-M images, the analyser was removed to account for all contributions. Temperature dependence measurements were performed at a 1 K/min heating rate by imaging a small area of the sample. Intensity was computed as the mean intensity of a subregion. Upon heating from SmC$_\text{P}^\text{H}$ phase, the SHG intensity strongly decreases. A weak signal still present in the SmC\textsubscript{AF} and SmA\textsubscript{AF}, completely disappears in the SmA phase. The SHG signal in the the SmC\textsubscript{AF} and SmA\textsubscript{AF} phases is more than 5 orders of magnitude smaller than in the SmC$_\text{P}^\text{H}$ phase. The SHG image in SmC\textsubscript{AF} phase was acquired at a lower scanning speed, using pixel binning (16X), higher camera gain, and increased pump laser power. The polarization of the incoming beam was selected along the cells' rubbing direction after confirmation that the maximum SHG signal is obtained in that case.

\section{Supplementary Figures}

\begin{figure}[H]
    \centering
    \includegraphics[width=1\linewidth]{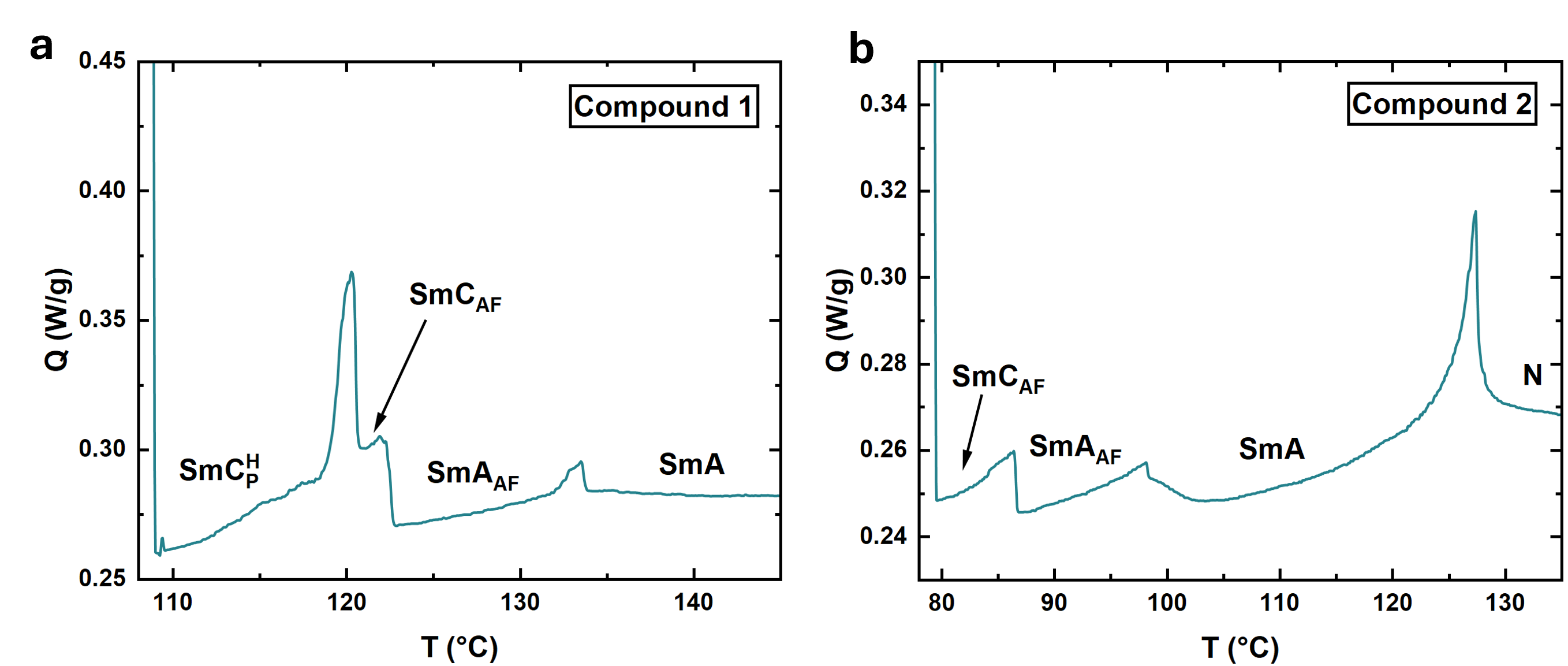}
    \caption{DSC cycles for a) compound \textbf{1} and b) compound \textbf{2}. These were recorded on cooling at 10$^\circ$C/min.}
    \label{fig:placeholder}
\end{figure}

\begin{figure}[H]
    \centering
    \includegraphics[width=1\linewidth]{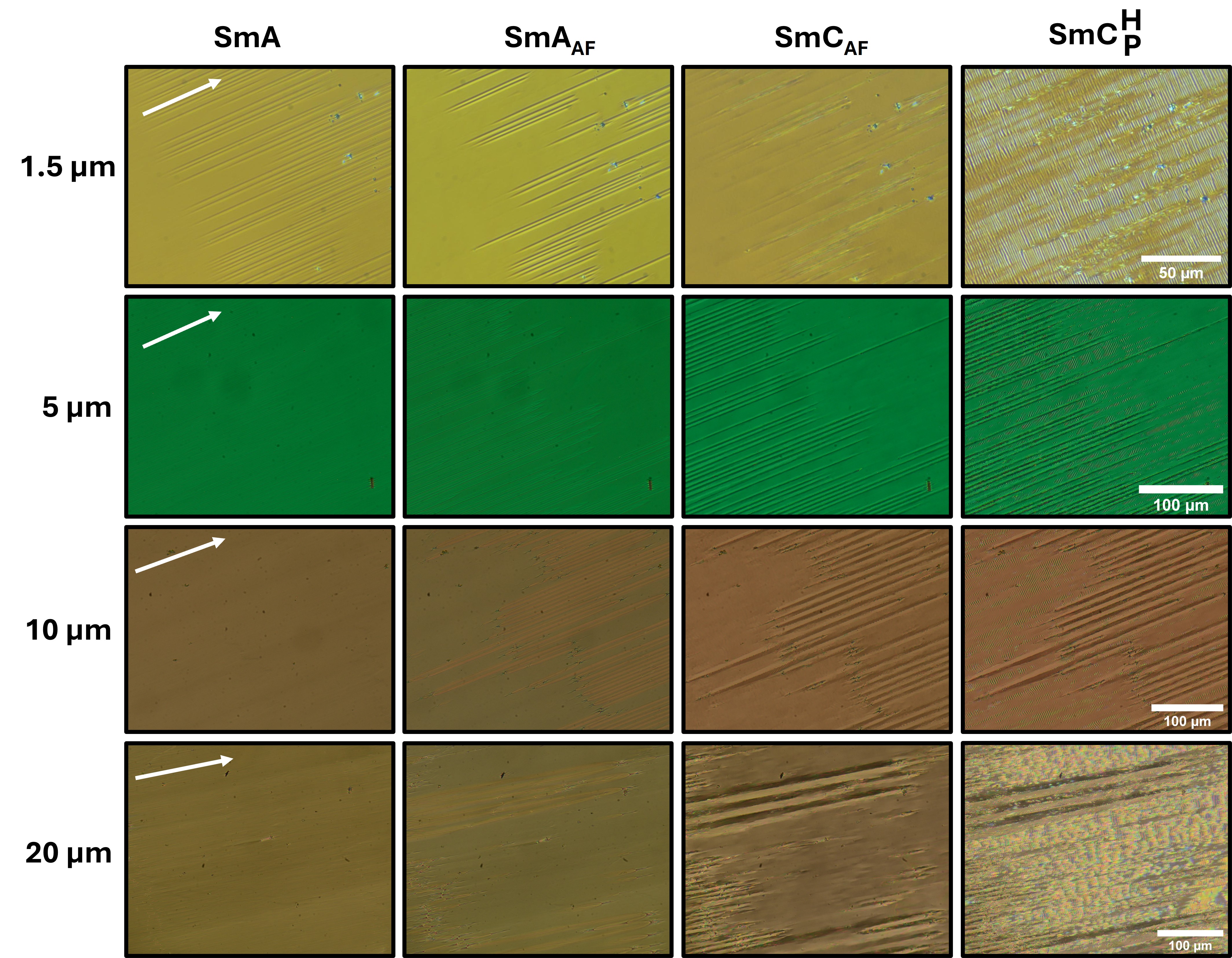}
    \caption{POM textures of compound \textbf{1} in planar aligned cells in with the cell thickness and in the phases marked.}
    \label{fig:placeholder}
\end{figure}

\begin{figure}[H]
    \centering
    \includegraphics[width=1\linewidth]{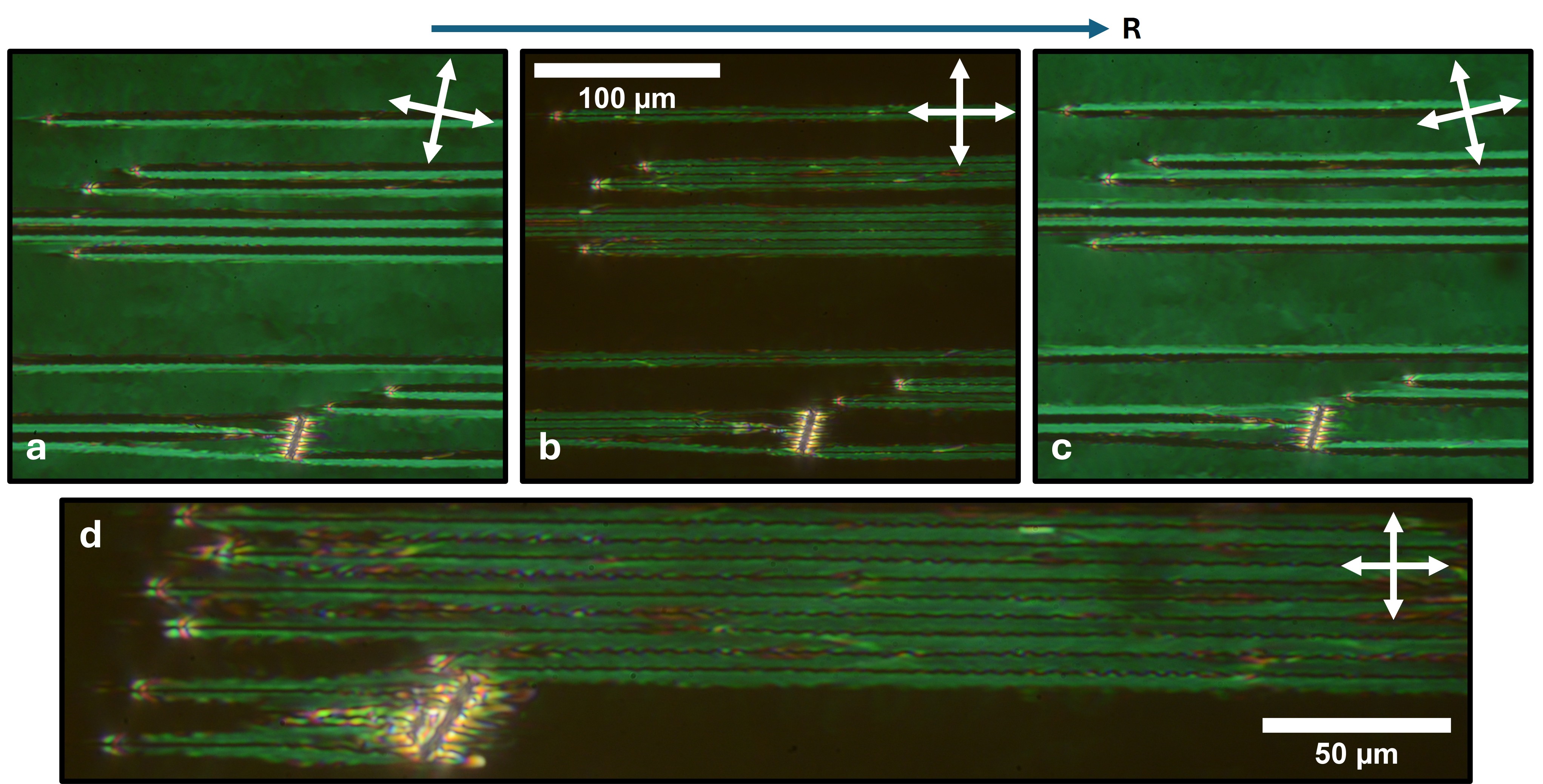}
    \caption{POM textures of the defects found in the SmC\textsubscript{AF} phase in a syn-parallel rubbed planar aligned 5 $\mu$m thick cell. These  defects are the evolution of the parabolic focal conic defects of the SmA phase \cite{1977_Rosenblatt}. The sample can be rotated to bring each half of the defect's sub domains into extinction indicating that each have has a different director orientation from themselves and the bulk uniform areas. The bulk uniform areas are brought into good extinction with the director fully parallel to the polariser orientation suggesting that the SmC\textsubscript{AF} phase is optically uniaxial or with the tilt plane perpendicular to the cell surfaces. Each sub-domain of the parabolic focal conic domain is joined by a discrimination line which appears to corkscrew.}
    \label{fig:placeholder}
\end{figure}

\begin{figure}[H]
    \centering
    \includegraphics[width=1\linewidth]{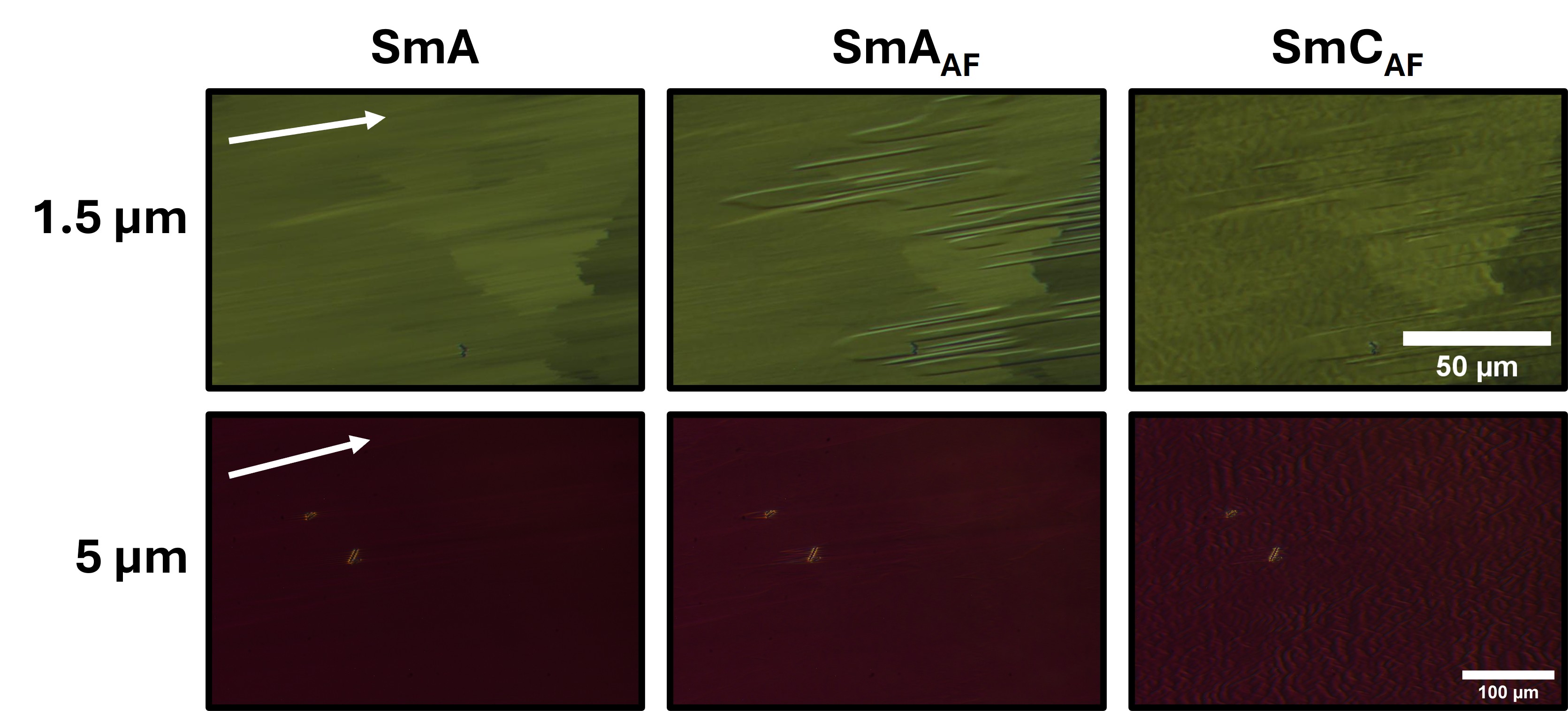}
    \caption{POM textures of compound \textbf{2} in planar aligned cells in with the cell thickness and in the phases marked. Chevron defects are observed in the SmA\textbf{AF} phase due to the reduction in the layer thickness. These chevrons are due to the smectic layer thickness. These chevrons do not appear to be due to the anti-ferroelectric domain shrinkage, (which does also occur) as has been observed in the N\textsubscript{S} phase \cite{2023_Chen}, as they have the wrong orientation, although they do seem to reduce in prominence throughout the SmC\textsubscript{AF} phase where the domain periodicity does increase.}
    \label{fig:placeholder}
\end{figure}

\begin{figure}[H]
    \centering
    \includegraphics[width=0.4\linewidth]{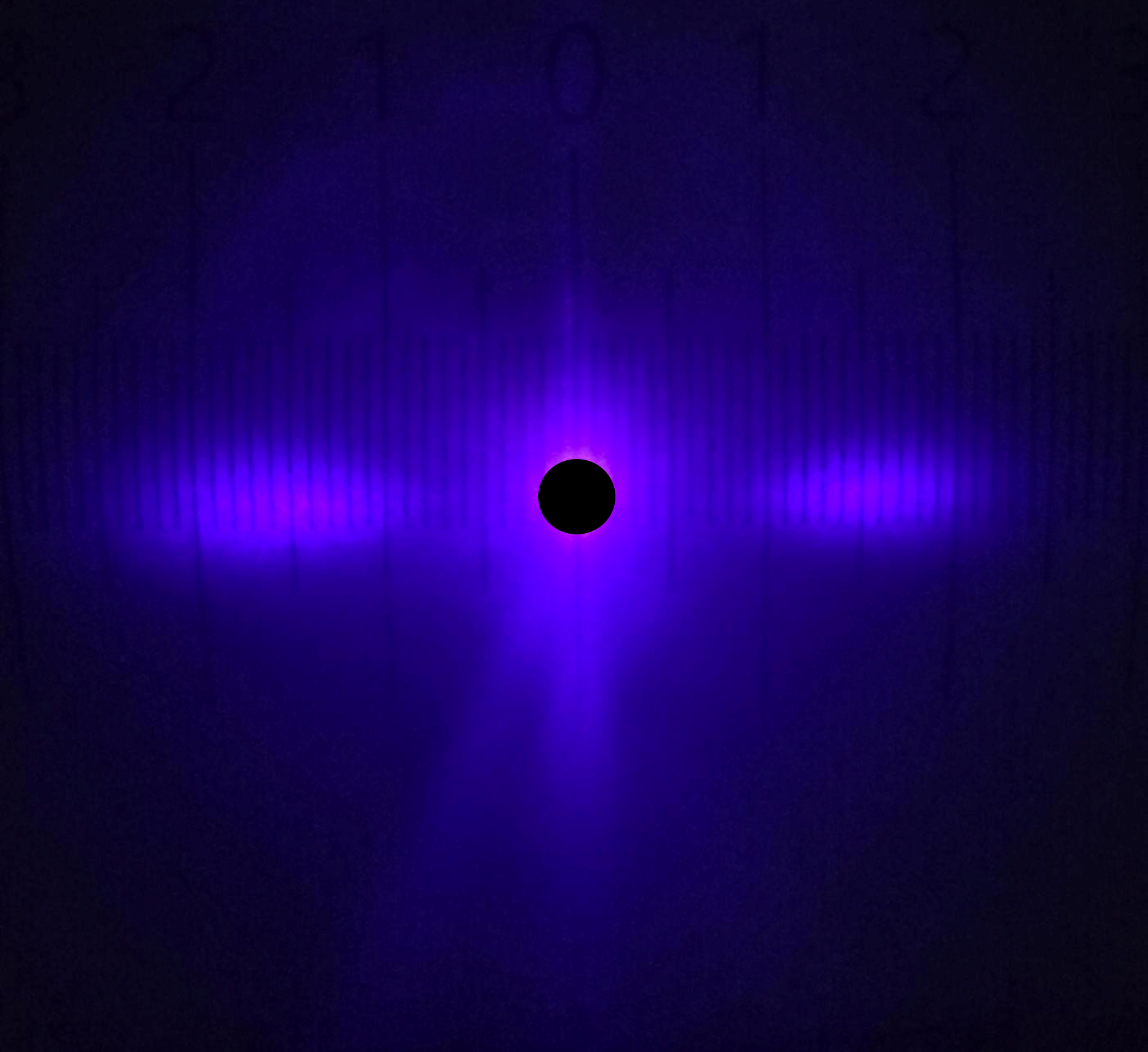}
    \caption{Light diffraction by compound \textbf{1} in the SmC$_\text{P}^\text{H}$ phase. This diffraction angle corresponds to a pitch length of 2 $\mu$m at 116$^\circ$C. This particular image was captured using a syn-parallel rubbed 5 $\mu$m planar aligned cell.}
    \label{fig:placeholder}
\end{figure}

\begin{figure}[H]
    \centering
    \includegraphics[width=1\linewidth]{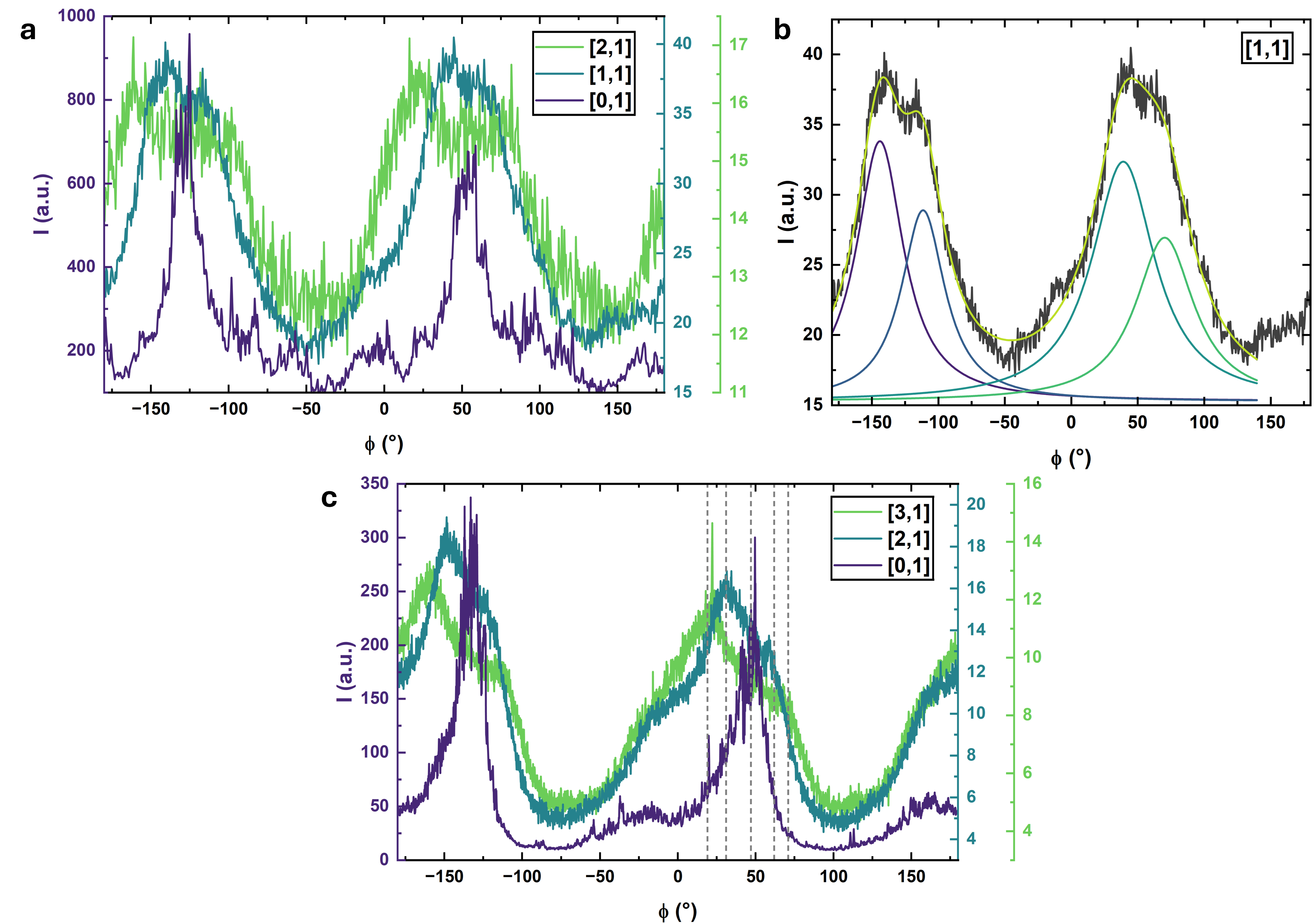}
    \caption{a) Azimuthal angular dependence of the X-ray scattering intensity for the various observed peaks for compound \textbf{1} in the SmA\textsubscript{AF} phase. b) The amount of spontaneous alignment allows the satellite peaks to be deconvoluted into two Lorentzians and thus obtain peak positions for each of the satellites. This is clearly not a monodomain sample as many other peaks can be observed and so a Lorentzian is used just as a more systematic way of obtaining the peak positions for this data than manual picking. c) Azimuthal angular dependence of the X-ray scattering intensity for compound \textbf{2} also in the SmA\textsubscript{AF} phase. Here the amount of spontaneous alignment is less and and so the peak positions were manually chosen and examples indicated by the grey dashed lines.}
    \label{fig:azimuthal}
\end{figure}

\begin{table}[H]
\centering
\caption{Angular positions for each of the peaks in fig. \ref{fig:azimuthal}a for compound \textbf{1}. The predicted values are obtained from equation (3).}
\begin{tabular}{|l|ll|ll|}
\hline
\textbf{[0,1]} & \multicolumn{1}{l|}{\textbf{[1,1]}} & \textbf{$\Delta\phi$} & \multicolumn{1}{l|}{\textbf{[2,1]}} & \textbf{$\Delta\phi$} \\ \hline
\multirow{2}{*}{-127.6} & \multicolumn{1}{l|}{-143.9} & 16.3 & \multicolumn{1}{l|}{-159.1} & 31.5 \\ \cline{2-5} 
 & \multicolumn{1}{l|}{-111.5} & -16.1 & \multicolumn{1}{l|}{-105.6} & -22.0 \\ \hline
\multirow{2}{*}{53.4} & \multicolumn{1}{l|}{39.1} & 14.3 & \multicolumn{1}{l|}{16.9} & 36.5 \\ \cline{2-5} 
 & \multicolumn{1}{l|}{70.2} & -16.8 & \multicolumn{1}{l|}{71.3} & -17.9 \\ \hline
\textbf{Average} & \multicolumn{2}{r|}{15.9} & \multicolumn{2}{r|}{27.1} \\ \hline
\textbf{Error} & \multicolumn{2}{r|}{0.9} & \multicolumn{2}{r|}{7.1} \\ \hline
\textbf{Predicted} & \multicolumn{2}{r|}{19.8} & \multicolumn{2}{r|}{35.8} \\ \hline
\end{tabular}
\end{table}

\begin{table}[H]
\centering
\caption{Angular positions for each of the peaks in fig. \ref{fig:azimuthal}c for compound \textbf{2}. The predicted values are obtained from equation (3).}
\begin{tabular}{|l|rr|rr|}
\hline
\textbf{[0,1]} & \multicolumn{1}{l|}{\textbf{[2,1]}} & \multicolumn{1}{l|}{\textbf{$\Delta\phi$}} & \multicolumn{1}{l|}{\textbf{[3,1]}} & \multicolumn{1}{l|}{\textbf{$\Delta\phi$}} \\ \hline
\multirow{2}{*}{-132.0} & \multicolumn{1}{r|}{-149} & 17 & \multicolumn{1}{r|}{-161} & -29 \\ \cline{2-5} 
 & \multicolumn{1}{r|}{-116} & -16 & \multicolumn{1}{r|}{-112} & 20 \\ \hline
\multirow{2}{*}{47.0} & \multicolumn{1}{r|}{31} & 16 & \multicolumn{1}{r|}{19} & -28 \\ \cline{2-5} 
 & \multicolumn{1}{r|}{62} & -15 & \multicolumn{1}{r|}{71} & 24 \\ \hline
\textbf{Average} & \multicolumn{2}{r|}{16.0} & \multicolumn{2}{r|}{25.3} \\ \hline
\textbf{Error} & \multicolumn{2}{r|}{$\pm$0.7} & \multicolumn{2}{r|}{$\pm$3.6} \\ \hline
\textbf{Predicted} & \multicolumn{2}{r|}{18.0} & \multicolumn{2}{r|}{26.0} \\ \hline
\end{tabular}
\end{table}

\begin{figure}[H]
    \centering
    \includegraphics[width=0.8\linewidth]{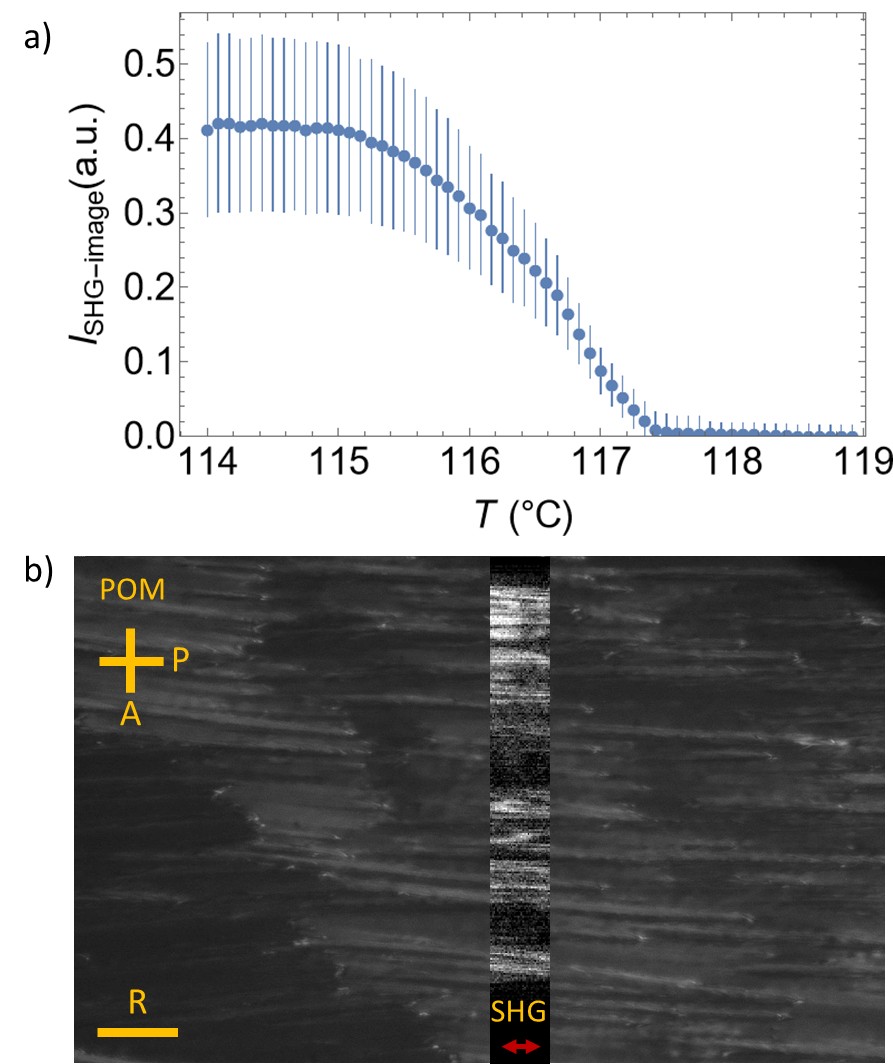}
    \caption{a) Average SHG image intensity for compound \textbf{1} in a 10 $\mu$m thick syn-parallel rubbed planar aligned cell on heating from the SmC$_\text{P}^\text{H}$ phase. Error bars denote the standard deviation. b) POM and SHG microscopy images of the SmC\textsubscript{AF} phase at 122$^\circ$C  in a 5 $\mu$m thick syn-parallel rubbed planar aligned cell showing the inhomogeneity of the very weak signal found there. P, A, and R denote polarizer, analyzer and rubbing directions, respectively. The red double arrow shows orientation of the IR pump laser polarization. }
    \label{fig:placeholder}
\end{figure}

\begin{figure}[H]
    \centering
    \includegraphics[width=1\linewidth]{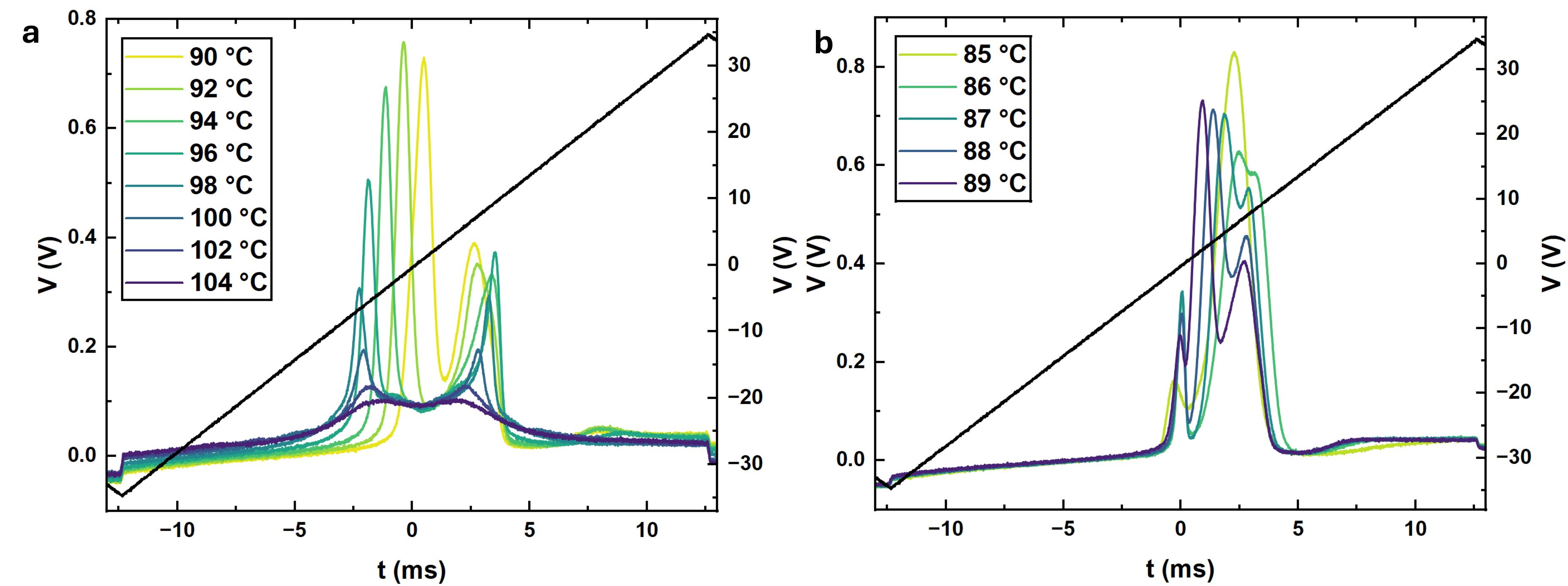}
    \caption{Current response measurements for compound \textbf{2} a) over the SmA-SmA\textsubscript{AF} phases and b) in the SmC\textsubscript{AF} phase. Data was recorded in a 5 $\mu$m thick cell with out of plane electrodes and no alignment layer.}
    \label{fig:placeholder}
\end{figure}

\begin{figure}[H]
    \centering
    \includegraphics[width=1\linewidth]{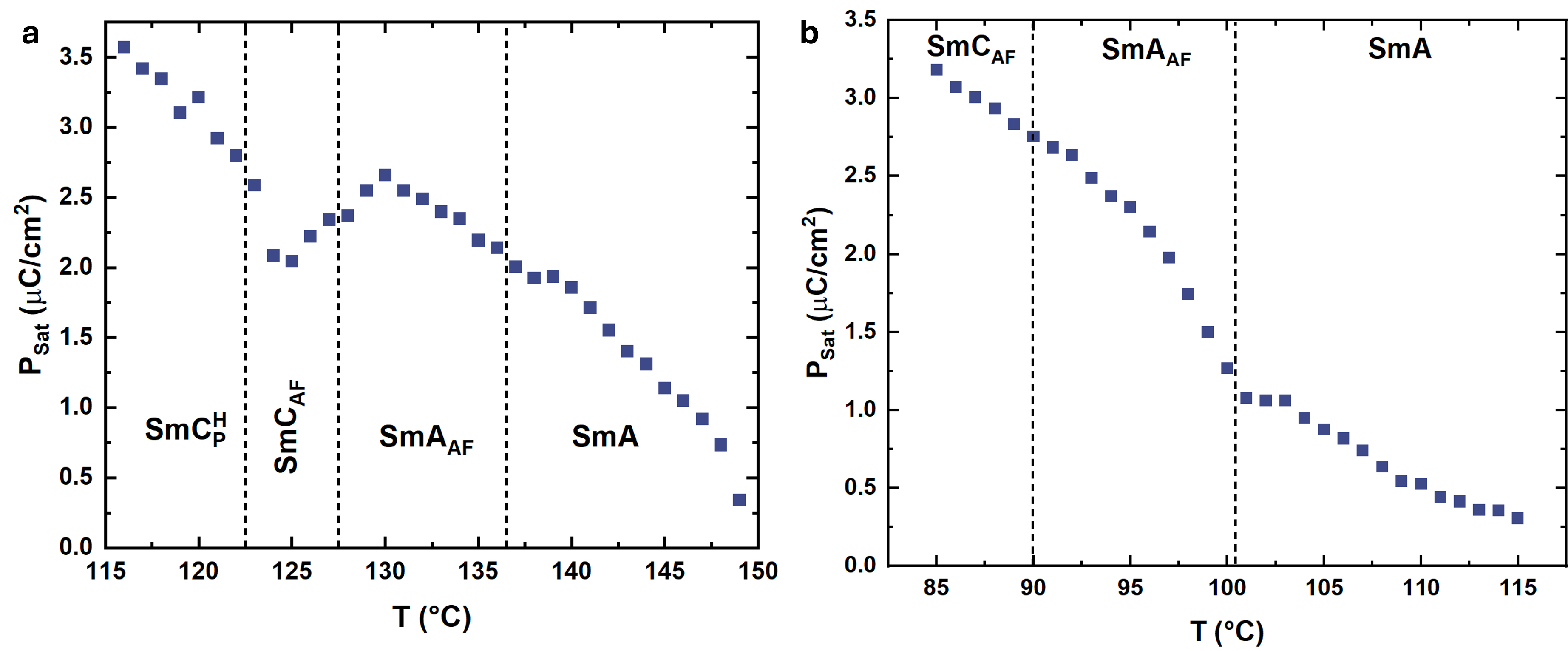}
    \caption{Measured saturated polarisation from the current responses for a) compound \textbf{1} and b) compound \textbf{2}}
    \label{fig:placeholder}
\end{figure}

\section{Organic Synthesis}

The total synthesis of compound \textbf{1} is outlined in \textbf{Scheme S1}. The synthesis of 4$^\prime$-(difluoro(3,4,5-trifluorophenoxy)methyl)-2,3$^\prime$,5$^\prime$-trifluoro-[1,1$^\prime$-biphenyl]-4-ol is described elsewhere \cite{2025_Hobbs2}. The preparation of compound 2 is also described elsewhere \cite{2024_Gibb}.

\begin{scheme}[H]
    \centering
    \includegraphics[width=1\linewidth]{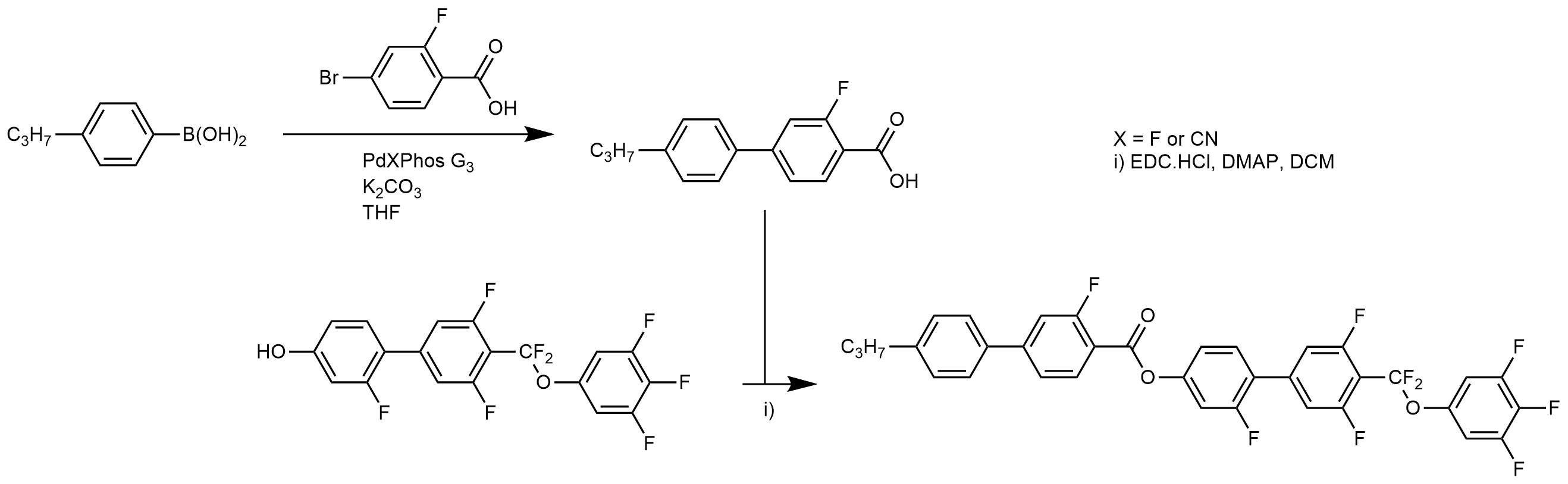}
    \caption{Synthesis of Compound 1}
    \label{fig:placeholder}
\end{scheme}

\subsection{Preparation of 4$^\prime$-propyl-3-fluorobiphenyl-4-carboxylic acid}

A reaction flask was charged with 4-propylphenyl boronic acid (5.41 g, 33 mmol) and dissolved in 200 mL of THF and 50 mL of 2M \ce{K2CO3}(aq). The resultant solution was sparged with \ce{N2}(g) for 30 mins. 4-bromo-2-fluorobenzoic acid (6.57 g, 30 mmol) was then added to the sparged solution which was then heated to 70$^\circ$C. Once at temperature, PdXPhos (G3) (0.5 mol\%) was then added in a single portion and the reaction left for 18 h with constant stirring. The reaction was then cooled, acidified using 2M HCl and extracted with EtOAc, the aqueous and organic layers separated, the organics being dried over \ce{MgSO4}. An off-white solid was retrieved under reduced pressure which was purified by re-crystallisation from MeCN to afford the title compound as a fine, white powder.

\begin{figure}[H]
    \centering
    \includegraphics[width=0.3\linewidth]{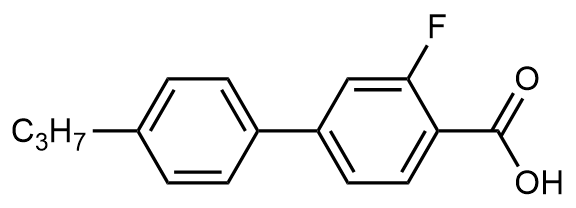}
    \caption*{\textit{4$^\prime$-propyl-3-fluorobiphenyl-4-carboxylic acid.}}
\end{figure}

\textbf{Yield:} \SI{7.1}{g} (91\%). 

\textbf{Rf (EtOAc):} 0.25.

\textbf{\ce{^1H} NMR} (\SI{501}{MHz}, DMSO) $\delta$ 13.21 (s, 1H, Ar-COO\textbf{H}), 7.92 (t, $J$ = 8.2 Hz, 1H, Ar-\textbf{H}), 7.69 (ddd, $J$ = 8.2, 2.2, 1.8 Hz, 2H, Ar-\textbf{H}), 7.65--7.58 (m, 2H, Ar-\textbf{H})\*, 7.31 (ddd, $J$ = 8.4, 2.0, 1.7 Hz, 2H, Ar-\textbf{H}), 2.61 (t, $J$ = 7.4 Hz, 2H, Ar-C\textbf{H$_2$}-CH$_2$), 1.62 (hept, $J$ = 7.2 Hz, 2H, CH$_2$-C\textbf{H$_2$}-CH$_3$), 0.91 (t, $J$ = 7.3 Hz, 3H, CH$_2$-C\textbf{H$_3$}). (\*Overlapping signals.)

\textbf{\ce{^{13}C\{^1H\}} NMR} (\SI{101}{MHz}, DMSO) $\delta$ 165.33 (d, $J$ = 3.2 Hz), 162.16 (d, $J$ = 256.8 Hz), 146.88 (d, $J$ = 8.7 Hz), 143.60, 135.46, 132.98, 129.56, 122.59 (d, $J$ = 3.2 Hz), 117.91 (d, $J$ = 10.3 Hz), 114.92 (d, $J$ = 23.3 Hz), 37.33, 24.40, 14.09.

\textbf{\ce{^{19}F} NMR} (\SI{376}{MHz}, DMSO) $\delta$ -109.81 (t, $J_{F-H}$ = 8.2 Hz, 1F, Ar-\textbf{F}).

\subsection{Preparation of Compound 1}
\sloppy A small vial was charged with 4$^\prime$-(difluoro(3,4,5-trifluorophenoxy)methyl)-2,3$^\prime$,5$^\prime$-trifluoro-[1,1$^\prime$-biphenyl]-4-ol (\SI{210}{mg}, \SI{0.5}{mmol}), 4$^\prime$-propyl-3-fluorobiphenyl-4-carboxylic acid (\SI{155}{mg}, \SI{0.6}{mmol}), EDC$\cdot$HCl (\SI{144}{mg}, \SI{0.75}{mmol}) and DMAP ($\sim$ 2~mol\%). Dichloromethane was added (conc. $\sim$ 0.1~M) and the suspension stirred until complete consumption of the phenol as judged by TLC. Once complete, the reaction solution was concentrated and purified by flash chromatography over silica gel with a gradient of Hexanes/DCM using a Combiflash NextGen300+ system. To remove ionic impurities, a pre-column containing a small amount of neutral alumina was used before samples entered the silica gel column. The chromatographed material was dissolved into the minimum quantity of DCM, filtered through a 0.2~\textmu m PTFE filter, concentrated to dryness and finally recrystallised from MeCN to afford the title compound as a white solid.

\begin{figure}[H]
    \centering
    \includegraphics[width=0.7\linewidth]{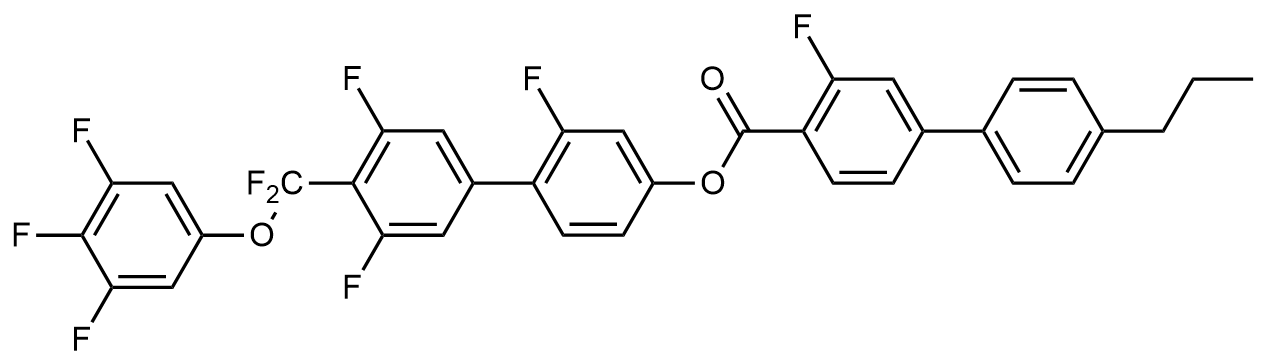}
    \caption*{4$^\prime$-(\textit{difluoro(3,4,5-trifluorophenoxy)methyl)-2,3$^\prime$,5$^\prime$-trifluoro-[1,1$^\prime$-biphenyl]-4-yl 3-fluoro-4$^\prime$-propyl-[1,1$^\prime$-biphenyl]-4-carboxylate}}
\end{figure}

\textbf{Yield:} \SI{238}{mg} (72\%). 

\textbf{Rf (DCM:Hexanes [1:1]):} 0.55.

\textbf{\ce{^1H} NMR} (\SI{400}{MHz}, CDCl$_3$) $\delta$ 8.15 (t, $J$ = 7.9 Hz, 1H, Ar-H), 7.57 (ddd, $J$ = 8.2, 2.0, 1.7 Hz, 2H, Ar-H), 7.55--7.49 (m, 2H, Ar-H)\*, 7.45 (dd, $J$ = 12.1, 1.7 Hz, 1H, Ar-H), 7.32 (ddd, $J$ = 8.3, 2.1, 2.1 Hz, 2H, Ar-H), 7.27--7.18 (m, 4H, Ar-H)\dag, 7.01 (dd, $J$ = 6.0, 1.7 Hz, 2H, Ar-H)\dag, 2.67 (t, $J$ = 7.4 Hz, 2H, Ar-CH$_2$-CH$_2$), 1.70 (hept, $J$ = 7.4 Hz, 2H, CH$_2$-CH$_2$-CH$_3$), 0.99 (t, $J$ = 7.3 Hz, 3H, CH$_2$-CH$_3$). (\textsuperscript{*} Overlapping peaks; \textsuperscript{\dag} Overlapping CDCl$_3$.)

\textbf{\ce{^{13}C\{^1H\}} NMR} (\SI{101}{MHz}, CDCl$_3$) $\delta$ 162.78 (d, $J$ = 261.8 Hz), 162.06 (d, $J$ = 4.2 Hz), 159.91 (dd, $J$ = 252.1, 5.7 Hz), 159.54 (d, $J$ = 252.4 Hz), 152.11 (d, $J$ = 11.3 Hz), 150.60 (ddd, $J$ = 251.0, 10.8, 5.0 Hz), 149.13 (d, $J$ = 8.9 Hz), 144.78--144.41 (m\textsubscript{apparent}), 144.10, 140.77 (t, $J$ = 11.1 Hz), 138.83 (dt, $J$ = 250.6, 14.9 Hz), 135.59, 132.97, 130.52 (d, $J$ = 3.9 Hz), 129.30, 127.03, 123.26 (d, $J$ = 13.1 Hz), 122.78, 122.50 (d, $J$ = 3.2 Hz), 120.14, 118.47 (d, $J$ = 3.6 Hz), 117.49, 115.50--115.02 (m)\*, 113.07 (dt, $J$ = 24.0, 3.2 Hz), 110.92 (d, $J$ = 25.8 Hz), 107.47 (d, $J$ = 23.9 Hz), 37.72, 24.45, 13.80. (\textsuperscript{*} Overlapping peaks.)

\textbf{\ce{^{19}F} NMR} (\SI{376}{MHz}, CDCl$_3$) $\delta$ -61.78 (t, $J_{F-F}$ = 26.3 Hz, 2F, -O-CF$_2$-Ar(F$_2$)), -107.23 (dd, $J_{F-H}$ = 12.5 Hz, $J_{F-H}$ = 7.5 Hz, 1F, Ar-F), -110.40 (td, $J_{F-F}$ = 26.3 Hz, $J_{F-H}$ = 10.9 Hz, 2F, Ar-F), -113.83 (t, $J_{F-H}$ = 9.8 Hz, 1F, Ar-F), -132.44 (dd, $J_{F-F}$ = 20.7 Hz, $J_{F-H}$ = 8.7 Hz, 2F, Ar-F), -163.12 (tt, $J_{F-F}$ = 20.9 Hz, $J_{F-H}$ = 5.5 Hz, 1F, Ar-F).

\begin{figure}[H]
\centering
\includegraphics[width=1\textwidth]{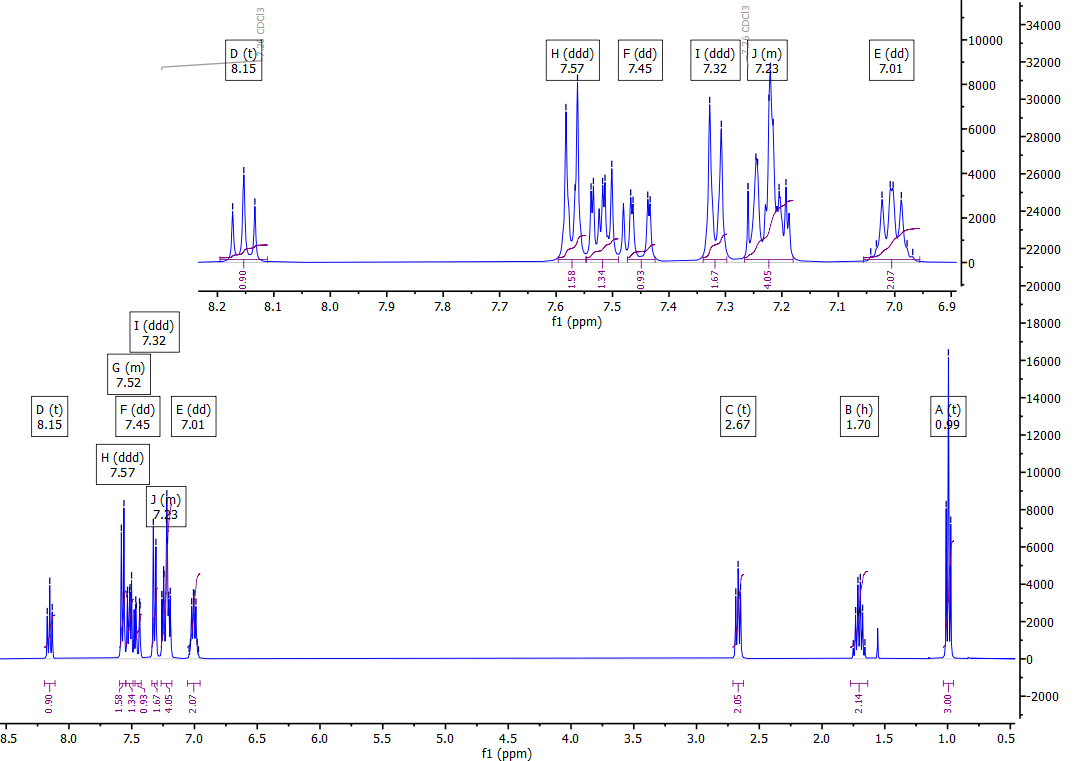}
\caption{\ce{^1H} NMR spectra of compound \textbf{1} dissolved in CDCl$_3$}
\end{figure}

\begin{figure}[H]
\centering
\includegraphics[width=1\textwidth]{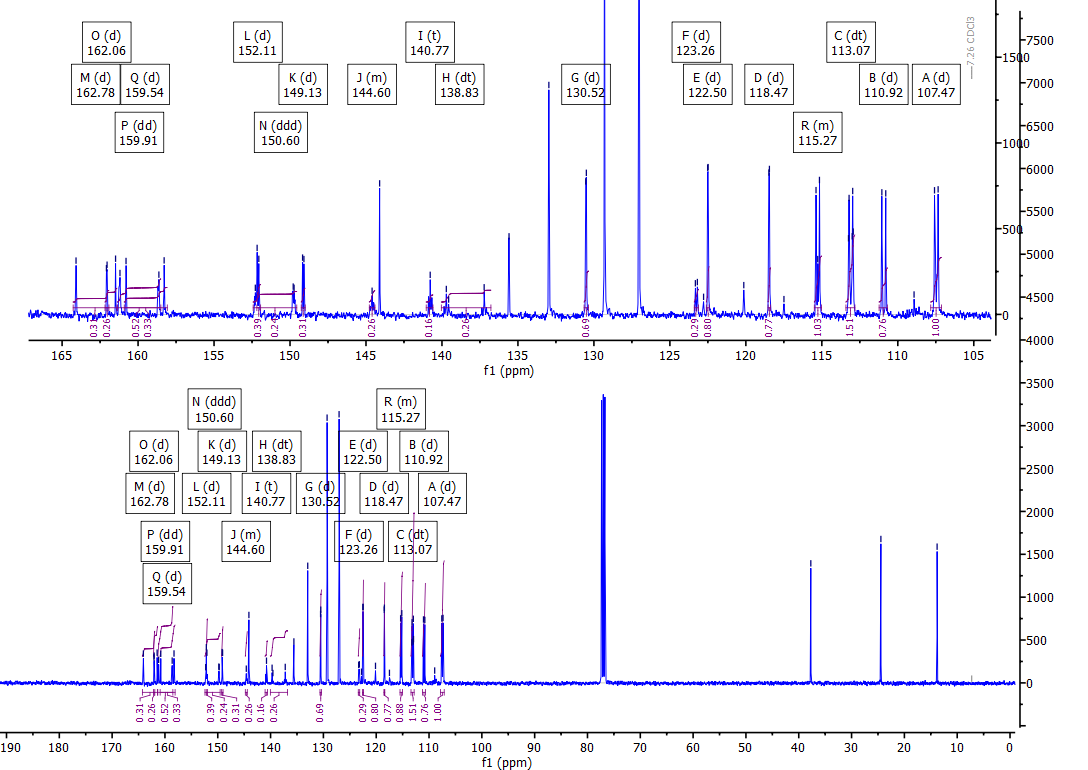}
\caption{\ce{^13C\{^1H\}} NMR spectra of compound \textbf{1} dissolved in CDCl$_3$}
\end{figure}

\begin{figure}[H]
\centering
\includegraphics[width=1\textwidth]{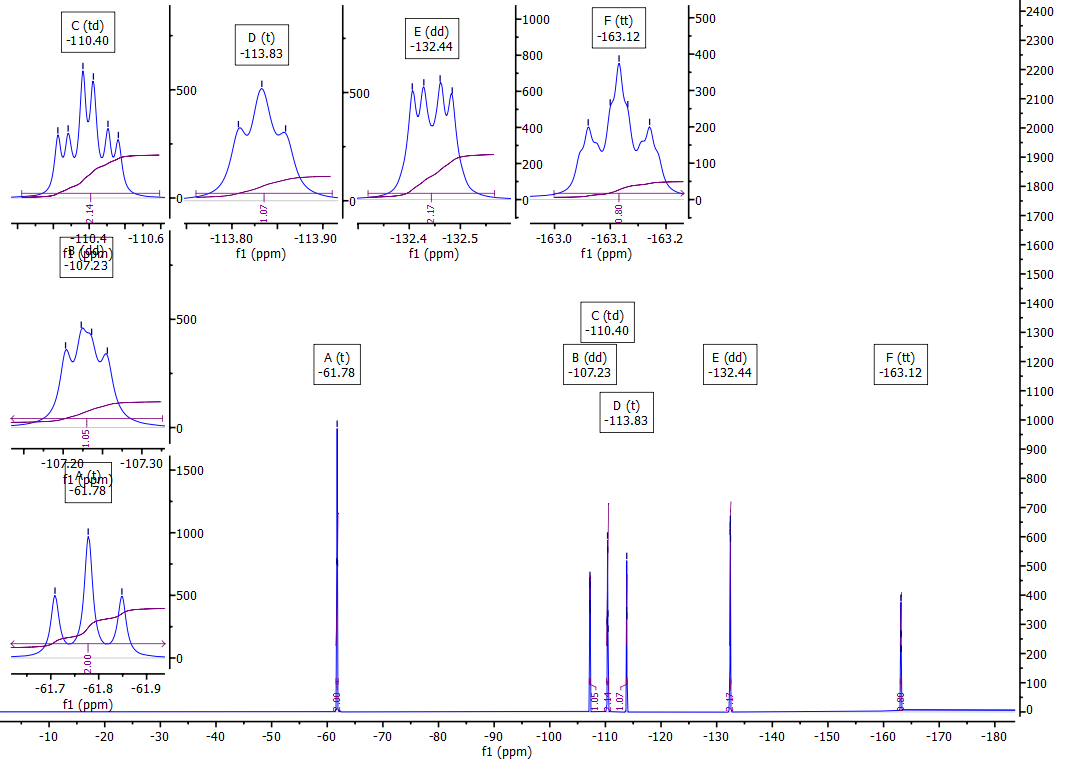}
\caption{\ce{^19F} NMR spectra of compound \textbf{1} dissolved in CDCl$_3$}
\end{figure}

\bibliographystyle{unsrt}
\bibliography{bib}